\documentclass[epj]{svjour}
\usepackage{latexsym}
\usepackage{graphics}
\usepackage{amsmath}
\usepackage{amsfonts}
\usepackage{slashed}
\usepackage{hyperref}

\begin{document}
\title{Revisiting neutrino masses in clockwork models}
\author{Aadarsh Singh\inst{1}% etc
\thanks{\emph{Email address:} aadarshsingh@iisc.ac.in}%
}                     % Do not remove
\institute{Indian Institute Of Science, \\
CV Raman Rd, Bengaluru, Karnataka 560012, India}
\date{Received: \today}
\abstract{
In this paper, we have studied a mechanism that naturally produces hierarchical masses using fine cancellation rather than widely used suppression mechanisms as in seesaw, clockwork or randomness-assisted localization models. We have also looked at various variants of the clockwork model and provided analytical expressions for their 0 modes. Few generalizations of clockwork models such as generalized CW and next to nearest neighbour interaction CW have already been explored by a few authors. All the variants of CW models have the same underlying principle i.e., suppression. In this study, it was found that non-local CW models relaxes the $\left| q \right| > 1$ constraint of ordinary CW models to produce localization. We also made a comparison among them and have shown that in some scenarios variants of CW are more efficient than ordinary CW. Additionally, the fine-cancellation (precision-prune) mechanism is depicted within the framework of the extra dimension picture. Finally, some phenomenological signatures of all these models are also discussed along with their benchmark points.
 % end of PACS codes
}%end of abstract
\maketitle
\section{Introduction}
\label{intro}
Neutrino masses are one of the most intriguing problems in flavour physics. To generate eV scale masses from natural weak scales of 100 GeV, it would require a suppression of $O(10^{12})$ in mass scale which seems unnatural. Several mechanisms have been proposed over the last 45 years starting with the seesaw mechanism and its variants Type II and Type III, etc. and then radiative mechanisms including the Zee and Scotogenic model etc. effective operators which generate neutrino masses \cite{minkowski1977mu},\cite{Yanagida:1979as},\cite{gell1979supergravity},\cite{glashow1980future},\cite{mohapatra1980neutrino},\cite{schechter1980neutrino}, \cite{de2016neutrino},\cite{Davidson:2008bu}. While the most natural mechanism of neutrino mass consists of Majorana neutrinos, in recent times enough attention has also been given to Dirac neutrino models. To explain the hierarchical nature of flavour parameters like masses $\&$ mixing angles, people have suggested localization of wavefunction approach such as in extra dimension models \cite{arkani2001neutrino},\cite{arkani2001electroweak},\cite{hill2001gauge},\cite{hallgren2005neutrino},\cite{randall1999alternative},\cite{randall1999large}. This wavefunction localization approach can be used to explain neutrino mass hierarchy as is done by several authors \cite{giudice2017clockwork},\cite{ibarra2018clockwork},\cite{hambye2017clockwork},\cite{park2018clockwork},\cite{Banerjee:2018grm},\\ \cite{Kitabayashi:2019qvi}. Equivalently localization can occur in theory space in a deconstruction picture.  That leads to the production of a highly suppressed Yukawa coupling and hence can be used to explain the required $O(10^{12})$ magnitude suppression in mass scale. Few authors have tried to explain clockwork structures using warped extra dimensions too that lead to a connection between clockwork and linear dilaton theory \cite{ahmed2017clockwork},\cite{park2018clockwork},\cite{giudice2017clockwork}. This clockwork approach has also been used by various authors to explain inflation, gravity, flavour mixing, axion, dark matter and various other shortcomings of SM \cite{teresi2017clockwork},\cite{hambye2017clockwork},\cite{wood2023clockwork},\cite{folgado2020gravity},\cite{niedermann2018higher},\cite{chiang2021testing},\cite{Kaplan:2015fuy}.

In this paper, firstly we have explored variants of clockwork models where the links in the theory space have been modified but the underlying suppression mechanism is kept intact. Some authors have already explored the generalized clockwork \cite{hong2019clockwork} and next-to-nearest neighbour clockwork \cite{ben2019generalized} and the both-sided clockwork scenario is similar to latticized extra dimension so results for these scenarios are known in the literature. We have modified these scenarios and applied them to fermions and found analytical expressions for the 0-mode eigenvector which for non-local theory spaces has a combinatorial factor in it. Since the factorial grows much faster than the exponential, the 0-mode localization can be bigger than the exponential localization such as in randomness-assisted models \cite{craig2018exponential}. However, for the models discussed here, the 0-mode components exhibit combinatorial rather than factorial dependence and, as a consequence, are unable to surpass the exponential. Also, we have described a fine-cancellation mechanism rather than the suppression mechanism to generate a hierarchical scale from the weak scale. This fine cancellation mechanism when implemented for neutrino mass generation, is capable of generating eV scale from TeV scale with natural orders for parameters. The couplings generated between SM and BSM fields are not tiny so these models are phenomenologically easily testable compared to suppression-based models. The cancellation mechanism is implemented in theory space that resembles the discretized extra dimension model.

The phenomenological signatures for models have been explored using the Flavor-changing neutral current (FCNC) of charged SM leptons. The SM with non-zero neutrino masses contribution for the branching ratio of $\mu \rightarrow e \gamma$ decay comes from one loop corrections and is significantly small $\approx$ $O(10^{-55})$. However, the introduction of these BSM-heavy neutral leptons drastically increases these BR numbers to within the grasp of current experimental bounds. The current most stringent bound for this decay comes from MEG with bound at BR $< 4.2 \times O(10^{-13})$ \cite{bao2016search} and can be used to restrict the parameter space of these models. We have given the benchmark points for these models producing the observed neutrino mass that survive the current MEG bounds but are within the reach of upcoming MEG-2 bounds BR $< 6 \times O(10^{-14})$ \cite{MEGII:2021fah}.

The paper begins by studying the clockwork variant models in section 2 $\&$ 3. In section 2, the model studied is both-sided clockwork (BCW) and it is shown that the suppression produced in this case is a few orders of magnitude stronger than CW for certain values of parameters. In section 3, non-local clockwork models are studied. The analytical expression of 0-mode for a few cases is given and it is shown that they contain a combinatorial term which for a certain range of parameters gives much bigger suppression than CW. In section 4, we revisited the neutrinos in Large extra dimension (LED) models from a dimensional deconstruction (DD) perspective. We have shown that localization of wavefunctions in this scenario is not a required condition to produce small masses and hence generated neutrino mass scales from TeV scales using only natural parameters. Finally, in section 5 a detailed analysis of BSM effects on FCNC BR is given and some comments about phenomenology in colliders and the contribution of these BSM fields to Higgs mass $\&$ width have been made.\\
Throughout the paper, the neutrino mass and mixing vlaues used are mentioned in Table \ref{tab:neutrino_mass_squared} and \ref{tab:mixing_angles}: \cite{ParticleDataGroup:2020ssz}, \cite{Esteban2022}

\begin{table}[h!]
\caption{Neutrino mass-squared differences for Normal and Inverted Hierarchy cases at $1\sigma$ range.}
\label{tab:neutrino_mass_squared}  % Give a unique label
\begin{tabular}{lll}
\hline\noalign{\smallskip}
\textbf{Hierarchy} & $\Delta m^2_{21}$ ($eV^2$) & $\Delta m^2_{32}$ ($eV^2$) \\
\noalign{\smallskip}\hline\noalign{\smallskip}
Normal Hierarchy & $7.39^{+0.21}_{-0.20} \times 10^{-5}$ & $2.449^{+0.032}_{-0.030} \times 10^{-3}$ \\
Inverted Hierarchy & $7.39^{+0.21}_{-0.20} \times 10^{-5}$ & $-2.509^{+0.032}_{-0.032} \times 10^{-3}$ \\
\noalign{\smallskip}\hline
\end{tabular}
\end{table}

\begin{table}[h!]
\caption{Values of $\theta_{12}$, $\theta_{23}$, and $\theta_{13}$ for Normal and Inverted Hierarchy cases at $1 \sigma$ range.}
\label{tab:mixing_angles}  % Give a unique label
\begin{tabular}{llll}
\hline\noalign{\smallskip}
\textbf{Hierarchy} & $\theta_{12}$ ($^\circ$) & $\theta_{23}$ ($^\circ$) & $\theta_{13}$ ($^\circ$) \\
\noalign{\smallskip}\hline\noalign{\smallskip}
Normal Hierarchy & $33.82^{+0.78}_{-0.76}$ & $48.3^{+1.2}_{-1.9}$ & $8.61^{+0.13}_{-0.13}$ \\
Inverted Hierarchy & $33.82^{+0.78}_{-0.76}$ & $48.6^{+1.1}_{-1.5}$ & $8.65^{+0.13}_{-0.12}$ \\
\noalign{\smallskip}\hline
\end{tabular}
\end{table}

\section{Local Clockwork Models - Mass Model }
The clockwork (CW) models use mechanisms based on suppression by heavy scales to obtain hierarchically small couplings, leading to mass scales that are much smaller than the fundamental parameters of the model. The full Lagrangian is described as
\begin{equation}
    \mathcal{L} = \mathcal{L}_{SM} + \mathcal{L}_{NP} + \mathcal{L}_{int}.
\end{equation} 
with $\mathcal{L}_{SM}$ representing the lagrangian for the standard model given by
\begin{equation}
    \mathcal{L}_{SM} = \mathcal{L}_{Gauge} + \mathcal{L}_{Fermion} + \mathcal{L}_{Higgs} + \mathcal{L}_{Yukawa}.
\end{equation} 
$ \mathcal{L}_{NP}$ represents the physics of new fields and $\mathcal{L}_{int}$ is the interaction lagrangian between SM fields and new fields. For the new physics lagrangian, we will consider chiral fermionic fields with only Dirac couplings.
\begin{equation}
\mathcal{L}_{NP} = \mathcal{L}_{kin} - \sum_{i,j=1}^{n} \overline{L_{i}}\mathcal{H}_{i,j}R_j + h.c. \label{1}
\end{equation}
with $\mathcal{H}_{i,j}$ being the Hamiltonian representing the connection between different chiral fields. $L_i$ and $R_i$ represent the left and right-handed clockwork gear fields respectively.

The Hamiltonian and new physics Lagrangian for the generalized CW 
 are given below \cite{hong2019clockwork}:-
\begin{equation}
    \mathcal{H}_{i,j} =m_i \delta_{i,j} + m_iq_i\delta_{i+1,j}.  
\end{equation}
\begin{equation} \mathcal{L}_{NP} = \mathcal{L}_{kin} - \sum_{i}^{n} m_i \overline{L_{i}}R_i - \sum_{i}^{n} m_iq_i \overline{L_{i}}R_{i+1}  + h.c. \label{GCW}
\end{equation}
with $m_i$ as the mass parameters and $q_i$ as the coupling strength parameters of the model. Corresponding to $\lambda_0$ = 0 eigenvalue mode, the eigenvector for the GCW is given by  
$$ \Lambda_0= \left\{(-1)^nq_nq_{n-1}\dots q_1,(-1)^{n-1}q_nq_{n-1}\dots q_2,\dots,(-1)^{1}q_n,1\right\} , $$
$$ \tilde{\Lambda}_0 = \mathcal{N}_0 \Lambda_0.$$
with $\mathcal{N}_0$ as the normalizing factor and $\tilde{\Lambda}_0$ is the normalized 0-mode. Hence for $q_i$ $>$ 1, there will be suppression in the components of 0-mode eigenvector as it will be localized on a certain site.
This case along with its various phenomenology has been explored by  \cite{hong2019clockwork}. The uniform CW (UCW) model is a limiting scenario of generalized CW with $q_i = q \hspace{0.2cm} \forall \hspace{0.2cm} i$ is studied in \cite{giudice2017clockwork},\cite{ibarra2018clockwork}.

\subsection{Both sided Clockwork}
\begin{figure}
\resizebox{0.5\textwidth}{!}{%
  \includegraphics{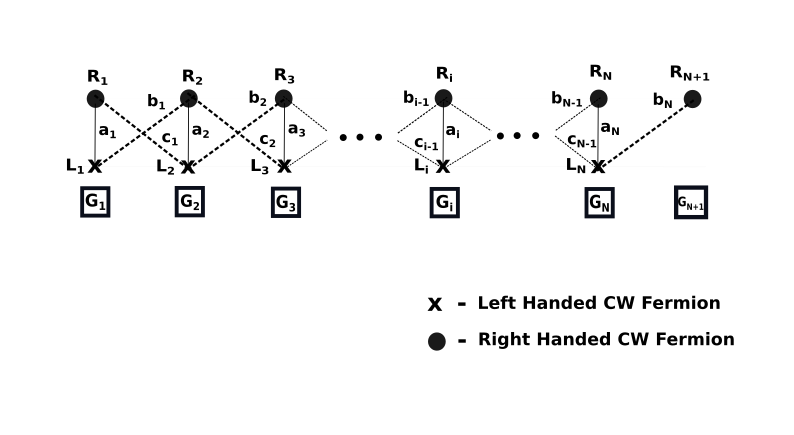}
}
    \caption{ CW with both-sided interactions. In CW notations, $a_i$ = $m_i$, $b_i$ = $m_iq_i$ and $c_i$ = $m_iq^{'}_i$.} \label{BCW}   
\end{figure}

This scenario is an extension of CW Hamiltonian where fermions of both chiralities are connected to each other for neighbouring matter fields. The Hamiltonian is diagrammatically represented in  figure \ref{BCW} and is written as
\begin{equation}
    \mathcal{H}_{i,j} =m_i \delta_{i,j} + m_iq_i\delta_{i+1,j}  + q'_im_i \delta_{i,j+1}. 
\end{equation}
\begin{equation}
\scalebox{0.8}{$
\mathcal{L}_{NP} = \mathcal{L}_{kin} 
- \sum_{i}^{n} m_i \overline{L_{i}}R_i 
- \sum_{i}^{n} m_i q_i \overline{L_{i}}R_{i+1} 
- \sum_{i}^{n-1} m_i q'_i \overline{L_{i+1}}R_i 
+ \text{h.c.}
$}
\label{bcw}
\end{equation}
with i $\in$ $\{1,2,...n\}$ and j $\in$ $\{1,2,...n+1\}$ and $q'$ also representing the coupling strength parameter similar to $q$ but coupling the chiral fields the other way. The Hamiltonian looks similar to the deconstruction scenario, but qualitatively it is quite different. In the deconstruction scenario, 0-mode was not always produced, certain relations among fundamental parameters have to be satisfied and also there were constraints on the size of the lattice for 0-mode to exist but in both-sided CW (BCW) all fundamental parameters can be independent of each other along with the arbitrary number of lattice sites, 0-mode is always produced as seen in the theorem in Appendix-\ref{app:linear_algebra}. Thus this gives us much more freedom from the constraints of fundamental parameters to work with.

 The matrix for fermionic mass in this model in  the basis $\{\overline{L}_1, \overline{L}_2, \ldots, \overline{L}_n\}$ and $\{R_1, R_2, \ldots, R_{n+1}\}$ is given by
\[
M_{BCW} = \left(
\begin{array}{cccccccc}
m_1 & q_1 m_1 & 0 & 0 & \dots & 0 & 0  \\
q'_{1} m_{1} & m_2 & q_2 m_2 & 0 & \dots & 0 & 0  \\
0 & q'_{2} m_{2} & m_3 & q_3 m_3 & \dots & 0 & 0  \\
\vdots & \vdots & \vdots & \ddots & \ddots & \vdots & \vdots  \\
0 & 0 & 0 & \dots & q'_{n-1} m_{n-1} & m_{n} & q_{n} m_{n}
\end{array} \right)_{n \times n+1}
\]

 Once again, the right-handed fermions will have a 0-mode as per the theorem. The null basis for right-fermionic field matrix $M^{\dagger}_{BCW}M_{BCW}$ is same as the null basis of $M_{BCW}$ as
 $$ M^{\dagger}_{BCW} M_{BCW}\Lambda_0 = M^{\dagger}_{BCW} (M_{BCW}\Lambda_0) = M^{\dagger}_{BCW} (\Vec{0}) = \Vec{0}$$
 In the uniform limit, the 0-mode eigenvector behaves as
 $$\left(\begin{array}{cccccccc} 1 & q & 0 & 0 & \dots & 0 & 0 & 0 \\
q^{'} & 1 & q & 0 & \dots & 0 & 0 & 0 \\
0 & q^{'} & 1 & q & \dots & 0 & 0 & 0 \\
\vdots & \vdots & \vdots & \ddots & \ddots & \vdots & \vdots & \vdots \\
0 & 0 & 0 & \dots & \dots & q^{'} & 1 & q\end{array}\right)\left(\begin{array}{c}v_1 \\ v_2 \\ v_3 \\ \vdots \\ v_{n+1}\end{array}\right)=\Vec{0}$$
This gives the following recurrence relation for the 0-mode component
$$ q'v_{i-1} + v_i + qv_{i+1} = 0, \hspace{1.5cm} i \in \{2,3,...n\} $$
with a boundary condition $v_1 = -qv_2 $. Solving the recurrence relation with boundary condition we get the $k^{th}$ component value as
$$ v_k = c\hspace{1mm} 2^{-k} \left(\left(-\frac{\sqrt{1-4 q q'}+1}{q}\right)^k-\left(\frac{\sqrt{1-4 q q'}-1}{q}\right)^k\right)$$
where $c$ is some arbitrary constant. It is fixed by imposing the normalization condition on the null vector $\sum_{i=1}^{n+1} v^*_i v_i$ = 1.
$$ c = \frac{1}{\sqrt{\sum_{i=1}^{n+1} \Bigl\{2^{-k} \left(\left(-\frac{\sqrt{1-4 q q'}+1}{q}\right)^k-\left(\frac{\sqrt{1-4 q q'}-1}{q}\right)^k\right)\Bigr\}^2 }} $$
 The eigenvalues for n = 2 case are given by  
% {\tiny
% \begin{multline}
%     \lambda_i =  \Bigl\{0,\frac{m^2}{2} \Bigl(-\sqrt{\left(-\left(q'_1\right)^2-q_1^2-q_2^2-2\right)^2-4 \left(q_1^2 \left(q'_1\right)^2-2 q_1 q'_1+q_1^2 q_2^2+q_2^2+1\right)}+\left(q'_1\right)^2+q_1^2 \nonumber \\  +q_2^2+2\Bigr),\frac{m^2}{2} \Bigr(\sqrt{\left(-\left(q'_1\right)^2-q_1^2-q_2^2-2\right)^2-4 \left(q_1^2 \left(q'_1\right)^2-2 q_1 q'_1 +q_1^2 q_2^2+q_2^2+1\right)}+\left(q'_1\right)^2 \nonumber \\  +q_1^2+q_2^2+2\Bigr)\Bigr\}
% \end{multline}
% }

\begin{multline}
    \scalebox{0.60}{ % Adjust the scale factor as needed
    $\lambda_i = \Bigl\{0, \frac{m^2}{2} \Bigl(-\sqrt{\left(-\left(q'_1\right)^2 - q_1^2 - q_2^2 - 2\right)^2 - 4 \left(q_1^2 \left(q'_1\right)^2 - 2 q_1 q'_1 + q_1^2 q_2^2 + q_2^2 + 1\right)} + \left(q'_1\right)^2 + q_1^2 + q_2^2 + 2\Bigr),$} \\
     \scalebox{0.65}{ $\frac{m^2}{2} \Bigl(\sqrt{\left(-\left(q'_1\right)^2 - q_1^2 - q_2^2 - 2\right)^2 - 4 \left(q_1^2 \left(q'_1\right)^2 - 2 q_1 q'_1 + q_1^2 q_2^2 + q_2^2 + 1\right)} + \left(q'_1\right)^2 + q_1^2 + q_2^2 + 2\Bigr)\Bigr\}$}
\end{multline}

and the unnormalized 0-mode eigenvector is given by
\begin{align}
    \Lambda_0 = \left\{-\frac{q_1 q_2}{q_1 q'_1-1},\frac{q_2}{q_1 q'_1-1},1\right\}
\end{align}
Hence as $q'_1$ $\rightarrow$ $\frac{1}{q_1}$, the 0-mode becomes more and more localized. In uniform limit, $q_i = q$ and $q'_i$ = $q'$ $\forall$ i , the eigenvalues and 0-mode eigenvector is
\begin{multline}
    \scalebox{0.70}{$ \lambda_i = \left\{0,-\frac{m^2}{2} \sqrt{4  (q'+q)^2+q'^4}+1+\frac{q'^2}{2}+q^2,\frac{m^2}{2} \sqrt{4 (q'+q)^2+q'^4}+1+\frac{q'^2}{2}+ q^2\right\} $} 
\end{multline}
\begin{align}
    \Lambda_0 = \left\{-\frac{q^2}{q' q-1},\frac{q}{q' q-1},1\right\} \nonumber
\end{align}
The first component of 0-mode will be greater than that in the normal CW model for $\lvert q'q$ - $1 \rvert$ $<$ $ 1 $.
As $q'$ $\rightarrow$ $\frac{1}{q}$, first two components $\rightarrow$ $\infty$ hence 0-mode localization on last site $\rightarrow$ 0. 

\begin{equation}
    (q|q')\in \mathbb{R}, 
    \left(q'<0\land \frac{2}{q'}<q<0\right)\lor \left(q'>0\land 0<q<\frac{2}{q'}\right)
\end{equation}
for these sets of values, the BCW model will produce greater suppression than ordinary CW. For n = 2, q = -3, CW produces $10^{-1}$ order suppression, but for this set of values with $q'$ = -0.33, this model produces $10^{-3}$ order of suppression, which is two orders of magnitude bigger suppression than CW. Similarly, the 0-mode eigenvector (not normalized) for n = 3 in the uniform limit can be found
\begin{align}
 \Lambda_0=&\left\{\frac{q^3}{2 q q'-1},-\frac{q^2}{2 q q'-1},-\frac{q \left(q q'-1\right)}{2 q q'-1},1\right\} \label{bcw-0-mode} 
\end{align}
again the first component of 0-mode will be greater than that in the normal CW model for $\lvert 2q'q$ - $1 \rvert$ $<$ $ 1 $. In this model apart from CW localization of $q^n$ in the numerator of the first component of 0-mode, there is a denominator too which depends on the extra parameter $q'$ introduced by coupling left-handed fermions of $i^{th}$ group to the right-handed fermions of $(i-1)^{th}$ group. The suppression of coupling produced will be more than CW in this scenario for the following set of parameters
\begin{align}
    (q|q')\in \mathbb{R}, 
    \left(q'<0\land \frac{1}{q'}<q<0\right)\lor \left(q'>0\land 0<q<\frac{1}{q'}\right) \nonumber
\end{align}
To compare with CW, we took n = 40 gears with q = -2 to produce eV mass from the TeV scale but here it can be done with n = 20 for q = -2 and $q'$ = -0.15. To get large localization, we took both $q$ and $q'$ to be of the same sign. For opposite signs, as can be seen from eq.(\ref{bcw-0-mode}), the denominator would not be that small and so localization will not be that large. This model will work better than CW for parameter values of $q'$ $\in$ $ [-0.539,0] $, there are three more intervals of smaller length but choosing $q'$ from those intervals will convert the natural hierarchy problem into the fine-tuning problem.

Now to study the effect of SM neutrino coupling with CW fermions perturbatively, we will consider the interaction term in Lagrangian of the same form as in GCW.
\begin{equation}
\mathcal{L}_{int}=-Y \widetilde{H} \bar{L}_{L} R_{n+1}+\text { h.c. } \label{lint}
\end{equation}
The Dirac mass matrix in the basis $\{\overline{\nu},\overline{L}_n, \overline{L}_{n-1},\hdots \overline{L}_1 \}$ and $\{R_{n+1}, R_n, R_{n-1},\hdots R_1 \}$ for Y = 0 is given by
\[ \Tilde{M} = m \begin{bmatrix}\makebox[1.5em][c]{0} & \makebox[1.5em][c]{0} & 0 & 0 & \dots & 0 \\
 \makebox[1.5em][c]{q} & \makebox[1.5em][c]{1} & q' & 0 & \dots & 0 \\
0 &  \makebox[1.5em][c]{q} & \makebox[1.5em][c]{1} & q' & \dots & 0 \\
\vdots & \vdots & \vdots & \ddots & \ddots & \vdots \\
0 & 0 & 0 & \dots & \makebox[1.5em][c]{q} & \makebox[1.5em][c]{1}
\end{bmatrix}_{(n+1) \times (n+1)}
\]
 The masses for right-handed CW fermions will be again given by $m\sqrt{\Tilde{\lambda_i}} $, with  $\Tilde{\lambda_i}$ denoting the eigenvalues of the matrix $\Tilde{M^\dagger}\Tilde{M}$/$m^2$.
 \begin{align*}
\scalebox{0.9}{$
 \widetilde{M^\dagger}\widetilde{M} = m^2 \begin{bmatrix}
    \makebox[1.5em][c]{$q^2$} & \makebox[1.5em][c]{$q$} & q q' & 0 & \dots & 0 \\
    \makebox[1.5em][c]{$q$} & \makebox[3em][c]{$1+q^2$} & q+q' & qq' & \dots & 0 \\
    qq' & \makebox[1.5em][c]{$q+q'$} & \makebox[4.2em][c]{$1+q^2+q'^2$} & q+q' & \dots & 0 \\
    \vdots & \vdots & \vdots & \ddots & \ddots & \vdots \\
    0 & 0 & 0 & \dots & \makebox[1.5em][c]{$q+q'$} & \makebox[3em][c]{$1+q'^2$}
\end{bmatrix}_{(n+1) \times (n+1)} 
$}
\end{align*}

For non-zero coupling Y, the eigenstates and eigenvalues change depending on the value of Y. Consider p to be the perturbative coupling strength then the Dirac mass matrix $\Tilde{M}$ becomes
\[ M = m \begin{bmatrix}\makebox[1.5em][c]{p} & \makebox[1.5em][c]{0} & 0 & 0 & \dots & 0 \\
 \makebox[1.5em][c]{q} & \makebox[1.5em][c]{1} & q' & 0 & \dots & 0 \\
0 &  \makebox[1.5em][c]{q} & \makebox[1.5em][c]{1} & q' & \dots & 0 \\
\vdots & \vdots & \vdots & \ddots & \ddots & \vdots \\
0 & 0 & 0 & \dots & \makebox[1.5em][c]{q} & \makebox[1.5em][c]{1}
\end{bmatrix}_{(n+1) \times (n+1)}
\]
The right-handed fermions have masses m$\sqrt{\lambda_i}$ with $\lambda_i$ being the eigenvalues of the following mass matrix:-
 \begin{align*}
\scalebox{0.9}{$
\frac{M^\dagger M}{m^2} =  \begin{bmatrix}
    \makebox[3em][c]{$p^2+q^2$} & \makebox[1.5em][c]{$q$} & q q' & 0 & \dots & 0 \\
    \makebox[1.5em][c]{$q$} & \makebox[3em][c]{$1+q^2$} & q+q' & qq' & \dots & 0 \\
    qq' & \makebox[1.5em][c]{$q+q'$} & \makebox[4.2em][c]{$1+q^2+q'^2$} & q+q' & \dots & 0 \\
    \vdots & \vdots & \vdots & \ddots & \ddots & \vdots \\
    0 & 0 & 0 & \dots & \makebox[1.5em][c]{$q+q'$} & \makebox[3em][c]{$1+q'^2$}
\end{bmatrix}_{(n+1) \times (n+1)}
$}
\end{align*}

Again we can write this matrix as the sum of the Dirac matrix without Yukawa neutrino coupling and a perturbation matrix.
$$ M^\dagger M = \Tilde{M^\dagger}\Tilde{M} + \delta M^2 $$
with perturbation matrix given by
\[ \delta M^2 = m^2 \begin{bmatrix}
    \makebox[1.5em][c]{$p^2$} & \makebox[3em][c]{$\mathbf{0}_{1\times n}$}  \\
    \makebox[3em][c]{$\mathbf{0}_{n\times 1}$} & \makebox[3em][c]{$\mathbf{0}_{n\times n}$}  \\
\end{bmatrix}_{(n+1) \times (n+1)} \]

As the perturbation matrix in CW fermions basis is again independent of neighbouring CW coupling strength parameters and depends on SM field coupling strength and the mass scale, the leading-order corrections to the eigenvalues are proportional to p.
$$
\delta \lambda_i=\left\langle \Lambda^{(i)}\left|\frac{\delta M^2}{m^2}\right| \Lambda^{(i)}\right\rangle=p^2 f(q,q')=O(p^2)
$$

$\Lambda^{(i)}$ denotes eigenvectors of the unperturbed matrix for BCW (both-sided clockwork) case eq.\eqref{bcw} and $f(q,q')$ denotes function f coming from the dependence of eigenvector $\Lambda_i's$ components on variables $q$ and $q'$. The leading order corrections are of the second order with respect to p. In this scenario, the perturbative eigenvector analysis in the appendix of \cite{hong2019clockwork} holds true up to the order of p.

\begin{figure}
\centering
\resizebox{0.3\textwidth}{!}{%
  \includegraphics{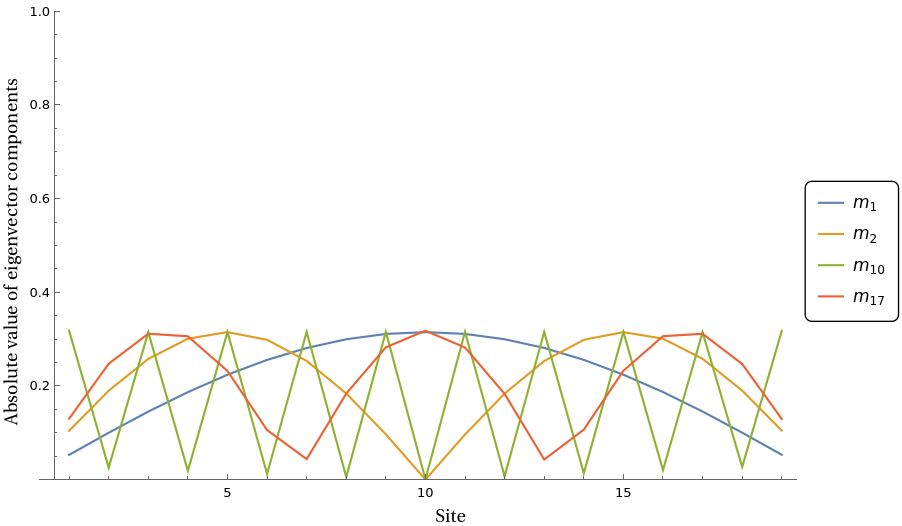} 
} \\
\resizebox{0.3\textwidth}{!}{%
  \includegraphics{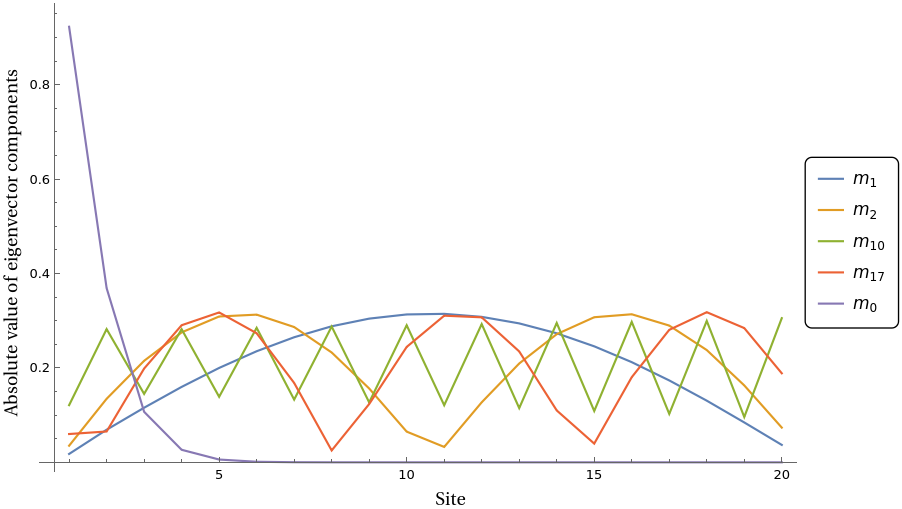}
}
    \caption{ The upper plot shows the absolute value of left-handed mass eigenvectors in terms of CW fields and the lower plot for right-handed mass eigenbasis before the Higgs achieves vev for n = 19 with $q = -2.5, q' = -0.11$. The localisation of the right-handed 0-mode with these parameters is evident from the plot for $m_0$ mode.} \label{BCW_local.}   
\end{figure}

\begin{figure}[h]
\centering
\resizebox{0.3\textwidth}{!}{%
  \includegraphics{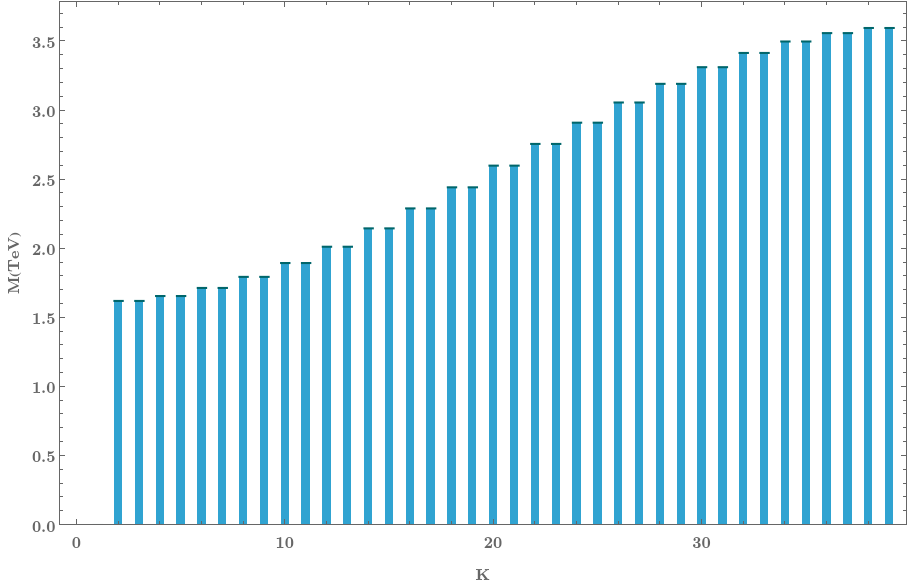} 
} \\
\resizebox{0.3\textwidth}{!}{%
  \includegraphics{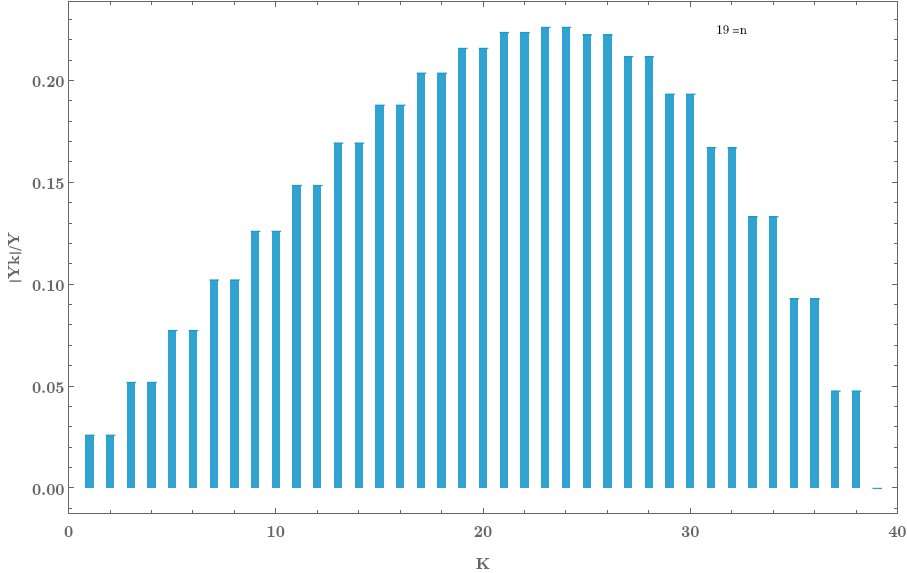}
}
     \caption{ The upper plot shows the mass distribution and the lower plot shows the coupling strength produced between SM neutrino and $K^{th}$ mass mode for the BCW scenario with parameters given in the Table \ref{tab:bp_points}.} \label{BCW_mass}
\end{figure}

Fig. \ref{BCW_local.} demonstrates the localisation of right-handed 0-mode and also delocalized eigenvectors in CW fields on different sites before Higgs achieves vev. The KK mass spectrum for Clockwork gears and their coupling strength produced between $K^{th}$ mass state and SM neutrino is shown in Fig. \ref{BCW_mass} with BP parameters.

\begin{table}
\caption{BP points for the model with $q_i = q, q'_i = q' \hspace{0.5cm} \forall i \in \{1, 2, \ldots, n \}$ producing neutrino mass in agreement with experimental data and avoiding current phenomenological constraints.}
\label{tab:bp_points} % Give a unique label
\centering
\begin{tabular}{lllll}  % Changed to five left-aligned columns
\hline\noalign{\smallskip}
\textbf{Hierarchy} & \textbf{n} & \textbf{$q$} & \textbf{$q'$} & \textbf{Yukawa Couplings} \\
\noalign{\smallskip}\hline\noalign{\smallskip}
Normal & 19 & $-5/2$ & $-0.11$ & $ \{y_1, y_2, y_3 \} = \{0.6, 1, 5 \}$ \\
\noalign{\smallskip}\hline\noalign{\smallskip}
Inverted & 19 & $-5/2$ & $-0.11$ & $ \{y_1, y_2, y_3 \} = \{4.92, 5, 1 \}$ \\
\noalign{\smallskip}\hline
\end{tabular} % with the correct table height (optional)
\end{table}
Table \ref{tab:bp_points} lists the benchmark points for the BCW model that are consistent with current phenomenological constraints. To compare, the standard CW suppression produced for $n=19$ and $q=-2.5$, is $\sim 10^{-8}$, while the BCW model with the specified parameters achieves a significantly stronger suppression of $\sim 10^{-14}$. Hence demonstrating the possibility of enhanced suppression with the BCW structure for the same site numbers.

\section{Non-local Clockwork Models}
\subsection{NNN CW}
In non-local CW theory space, 
 matter fields corresponding to groups which are not adjacent in the theory space/moose diagram also have link fields connecting them. These connections are formulated in the model by modifying the underlying Hamiltonian in the Lagrangian of the model. The Hamiltonian for NNN-CW (next-to-nearest neighbour clockwork) \cite{ben2019generalized} is given by 
\begin{align}
    \mathcal{H}_{i,j} =m_i \delta_{i,j} + q_i^{(1)}m_i\delta_{i+1,j} +q_i^{(2)}\delta_{i+2,j} \label{NNN} 
\end{align}
\begin{align} \mathcal{L}_{NP} = & \mathcal{L}_{kin} - \sum_{i}^{n} m_i \overline{L_{i}}R_i - \sum_{i}^{n} m_iq_i^{(1)} \overline{L_{i}}R_{i+1}  \nonumber \\  & -\sum_{i}^{n-1} m_iq_i^{(2)} \overline{L_{i}}R_{i+2} + h.c.
\end{align}
with i $\in$ $\{1,2,...n\}$ and j $\in$ $\{1,2,...n+1\}$.

\begin{figure}
\resizebox{0.5\textwidth}{!}{%
  \includegraphics{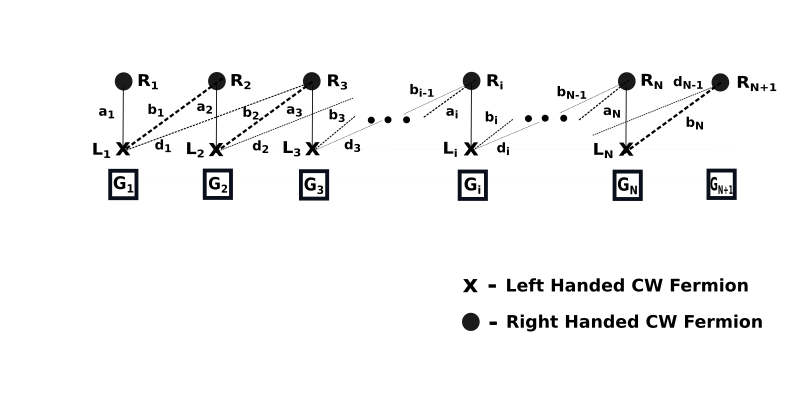}
}
    \caption{ CW with NNN (Next to Nearest Neighbour) interactions. In CW notations, $a_i$ = $m_i$, $b_i$ = $m_iq_i^{(1)}$ and $d_i$ = $m_iq_i^{(2)}$. }
\end{figure}

 In this model too, the mass matrix for right-handed CW fermions will have at least one 0 mode as per the theorem in Appendix-\ref{app:linear_algebra}. The matrix for the above Hamiltonian eq.\eqref{NNN} in $\{\overline{L}_1,\overline{L}_2,...\overline{L}_n \} $
 and  $\{R_1,R_2,....R_{n+1} \}$ basis is given by 
\[ M_{CW} = \begin{bmatrix}
    \makebox[1.5em][c]{$m_1$} & \makebox[2em][c]{$m_1q_1^{(1)}$} & \makebox[3em][c]{$m_1q_1^{(2)}$} & 0 & \dots & 0 \\
    \makebox[1.5em][c]{0} & \makebox[1.5em][c]{$m_2$} & \makebox[2em][c]{$m_2q_2^{(1)}$} & \makebox[1.5em][c]{$m_2q_2^{(2)}$} & \dots & 0 \\
    0 & \makebox[1.5em][c]{0} & \makebox[1.5em][c]{$m_3$} & \makebox[2em][c]{$m_3q_3^{(1)}$} & \dots & 0 \\
    \vdots & \vdots & \vdots & \ddots & \ddots & \vdots \\
    0 & 0 & \dots & \makebox[3em][c]{$m_{n-1}$} & \makebox[3.7em][c]{$m_{n-1}q_{n-1}^{(1)}$} & \makebox[4.8em][c]{$m_{n-1}q_{n-1}^{(2)}$} \\
    0 & 0 & 0 & \dots & \makebox[1.5em][c]{$m_{n}$} & \makebox[2em][c]{$m_nq_n^{(1)}$}
\end{bmatrix}_{n \times n+1} \]
The $K^{th}$ component for null space basis of NNN CW in the uniform limit case, $m_i$ = m, $q_i^{(1)} = q$ and $q_i^{(2)}$ = $q'$ $\forall$ i is given by
\begin{align}
   \Lambda_0^K = \sum_{\{k_i,k_j\}}^{}&\frac{(k_i+k_j)!}{k_i!k_j!}\frac{(-mq')^{k_i}(-mq)^{k_j}}{m^{k_i+k_j}}   \\
   with \hspace{1.5cm} 2k_i + k_j& =  n + 1 - K \\
   \Lambda_0 = \{\Lambda_0^1,&\Lambda_0^2,\hdots,\Lambda_0^{n+1}\} \nonumber
\end{align}
$k_i$ and $k_j$ $\in$ $\mathbb{N}_0 = \{0,1,2,3,4, ... \}.$ The normalized 0-mode $\Tilde{\Lambda}_0$ is given by $\mathcal{N}_0\Lambda_0$, with $\mathcal{N}_0$ representing the normalizing
factor. Using $\Tilde{\Lambda}_0^\dagger \Tilde{\Lambda}_0$ = 1, we get
\[ \mathcal{N}_0 = \frac{1}{\sqrt{\sum_{i=1}^{n+1}\Lambda_0^{i*} \Lambda_0^{i}}} \]

The 0-mode eigenvector for n = 2 case is 
\begin{align}
   \Lambda_0 =  \left\{-\frac{m_{2} m_1q_1^{(2)}-m_1q_1^{(1)} m_2q_2^{(1)}}{m_1 m_2},-\frac{m_2q_2^{(1)}}{m_2},1\right\}
\end{align}
in the uniform limit, $m_i$ = m, $m_iq_i^{(1)}$ = mq and $m_iq_i^{(2)}$ = $q'$ $\forall$ i, it will reduce to
\begin{align}
    \Lambda_0 = \left\{-(q'-q^2),-q,1\right\} \nonumber
\end{align}
it will produce bigger suppression than CW for 
\begin{align}
    (q|q')\in \mathbb{R}, 
   & \left( q<0\land \left(q'\leq 0\lor q'\geq 2 q^2\right)\right)\lor q=0\lor \nonumber \\ &\left(0<q\land \left(q'\leq 0\lor q'\geq 2 q^2\right)\right)
\end{align}
Now for n = 3 in the limiting case, the 0-mode eigenvector is
\begin{align}
    \left\{-(q^3-2  q q'),-(q'-q^2),-q,1\right\} \label{q-ve}
\end{align}
with conditions 
\begin{align}
    (q|q')\in \mathbb{R},
   & \left(q<0\land \left(q'<0\lor q'>q^2-q^3\right)\right)\lor \nonumber \\ & \left(0<q\land \left(q'<0\lor q'>q^3+q^2\right)\right) 
\end{align}
it produces greater suppression than CW.

\begin{figure}
\centering
\resizebox{0.3\textwidth}{!}{%
  \includegraphics{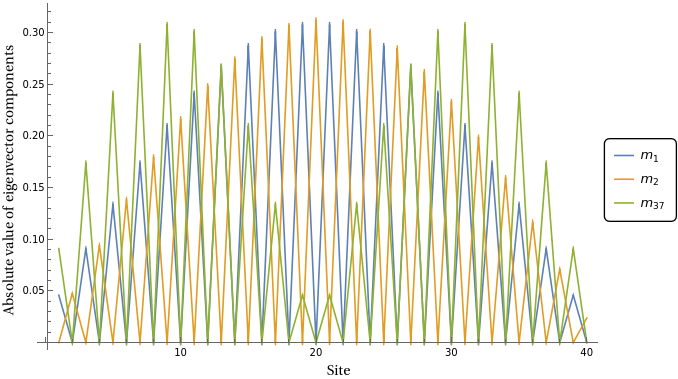} 
} \\
\resizebox{0.3\textwidth}{!}{%
  \includegraphics{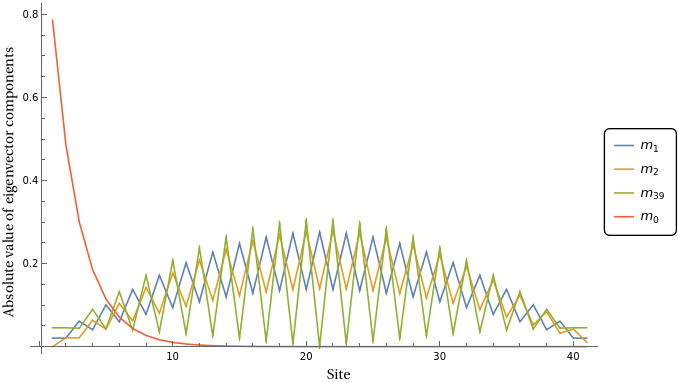}
}
    \caption{ The upper plot shows the absolute value of left-handed mass eigenvectors in terms of CW fields and the lower plot for right-handed mass eigenbasis before the Higgs achieves vev for n = 40 with $q = q' = -1$. The localisation of the 0-mode with these parameters is evident from the plot for $m_0$ mode.} \label{NNN_local}
\end{figure}

To compare with CW, we took n = 40 gears with m = 1 TeV, q = -2 to produce O(1) eV mass from the TeV scale, here it can be done with n = 25 for q = -2 and $q'$ = -3. One will notice that we took $-ve$ values for coupling parameters $q_i$, this is necessary to get bigger components in 0-mode, for $+ve$ values of $q_i$, the components cancel within themselves as can be seen in eq.\eqref{q-ve} and hence does not give large localization. Apart from faster localization, the NNN CW model relaxes the condition of $q > 1$ for localization to take place. This model achieves localization even for $|q|$ = 1 due to extra combinatorics factors in the components of 0-mode. This is explicitly shown in Fig. \ref{NNN_local} and compared with other models in Table \ref{tab:clockwork_model_comparison}.
Again to study the effect of SM neutrino coupling with NNN CW fermions perturbatively, we will consider the interaction term in Lagrangian of the same form as before eq.\eqref{lint}.
\begin{equation}
\mathcal{L}_{int}=-Y \widetilde{H} \bar{L}_{L} R_{n+1}+\text { h.c. }
\end{equation}
The Dirac mass matrix in the basis $\{\overline{\nu},\overline{L}_n, \overline{L}_{n-1},\hdots \overline{L}_1 \}$ and $\{R_{n+1}, R_n, R_{n-1},\hdots R_1 \}$ for Y = 0 is given by
\[ \Tilde{M} = m \begin{bmatrix}\makebox[1.5em][c]{0} & \makebox[1.5em][c]{0} & 0 & 0 & \dots & 0 \\
 \makebox[1.5em][c]{q} & \makebox[1.5em][c]{1} & 0 & 0 & \dots & 0 \\
q' &  \makebox[1.5em][c]{q} & \makebox[1.5em][c]{1} & 0 & \dots & 0 \\
0 & q' & q & 1 & \hdots & 0 \\
\vdots & \vdots & \vdots & \ddots & \ddots & \vdots \\
0 & 0 & 0 & \dots & \makebox[1.5em][c]{q} & \makebox[1.5em][c]{1}
\end{bmatrix}_{(n+1) \times (n+1)}
\]
 The masses for right-handed CW fermions will be again given by $m\sqrt{\Tilde{\lambda_i}} $, with  $\Tilde{\lambda_i}$ denoting the eigenvalues of the matrix $\Tilde{M^\dagger}\Tilde{M}$/$m^2$.
 \begin{align*}
\scalebox{0.7}{$
\Tilde{M^\dagger}\Tilde{M} = m^2 \begin{bmatrix}
q^2+q'^2 & q+qq' &  q' & 0 & \dots & 0 & 0\\
q+qq' & 1+q^2+q'^2 & q+qq' & q' & \dots & 0 & 0\\
q' & q+qq' & 1+q^2+q'^2 & q+qq' & \dots & 0 & 0\\
\vdots & \vdots & \vdots & \ddots & \ddots & \vdots & \vdots\\
0 & 0 & \dots &q'& q+qq' & 1+q^2 & q \\
0 & 0  & \dots & 0 & q' & q & 1
\end{bmatrix}_{(n+1) \times (n+1)}
$}
\end{align*}
For non-zero coupling Y, the eigenstates and eigenvalues change depending on the value of Y. Consider p to be the perturbative coupling strength then the Dirac mass matrix $\Tilde{M}$ becomes
\[ M = m \begin{bmatrix}\makebox[1.5em][c]{p} & \makebox[1.5em][c]{0} & 0 & 0 & \dots & 0 \\
 \makebox[1.5em][c]{q} & \makebox[1.5em][c]{1} & 0 & 0 & \dots & 0 \\
q' &  \makebox[1.5em][c]{q} & \makebox[1.5em][c]{1} & 0 & \dots & 0 \\
0 & q' & q & 1 & \hdots & 0 \\
\vdots & \vdots & \vdots & \ddots & \ddots & \vdots \\
0 & 0 & 0 & \dots & \makebox[1.5em][c]{q} & \makebox[1.5em][c]{1}
\end{bmatrix}_{(n+1) \times (n+1)}
\]
The right-handed fermions have masses m$\sqrt{\lambda_i}$ with $\lambda_i$ being the eigenvalues of the following mass matrix:-

{\tiny
\[ \frac{M^\dagger M}{m^2} =  \begin{bmatrix}
p^2+q^2+q'^2 & q+qq' &  q' & 0 & \dots & 0 & 0\\
q+qq' & 1+q^2+q'^2 & q+qq' & q' & \dots & 0 & 0\\
q' & q+qq' & 1+q^2+q'^2 & q+qq' & \dots & 0 & 0\\
\vdots & \vdots & \vdots & \ddots & \ddots & \vdots & \vdots\\
0 & 0 & \dots &q'& q+qq' & 1+q^2 & q \\
0 & 0  & \dots & 0 & q' & q & 1
\end{bmatrix}_{(n+1) \times (n+1)}
\]
}
Again we can write this matrix as the sum of the Dirac matrix with 0 neutrino coupling and a perturbation matrix.
$$ M^\dagger M = \Tilde{M^\dagger}\Tilde{M} + \delta M^2 $$
with the perturbation matrix given by
\[ \delta M^2 = m^2 \begin{bmatrix}
p^2 & \mathbf{0}_{1\times n}  \\
\mathbf{0}_{n\times 1} & \mathbf{0}_{n\times n}
\end{bmatrix}_{(n+1) \times (n+1)}
\]

As the perturbation matrix in CW fermions basis is again independent of neighbouring CW coupling strength parameters and depends on SM field coupling strength, the leading-order corrections to the eigenvalues are proportional to p.
$$
\delta \lambda_i=\left\langle \Lambda^{(i)}\left|\frac{\delta M^2}{m^2}\right| \Lambda^{(i)}\right\rangle=p^2 f(q,q')=O(p^2)
$$
again $\Lambda_i$ denotes eigenvectors of unperturbed matrix for next-to-nearest neighbour clockwork (NNN-CW) case and $f(q,q')$ denotes function f coming from dependence of eigenvector $\Lambda_i's$ components on variables q and $q'$. Once again the leading order corrections are of the second order with respect to p. In this scenario too, the perturbative eigenvector analysis in the appendix of \cite{hong2019clockwork} holds true up to the order of p.

\begin{figure}
\centering
\resizebox{0.3\textwidth}{!}{%
  \includegraphics{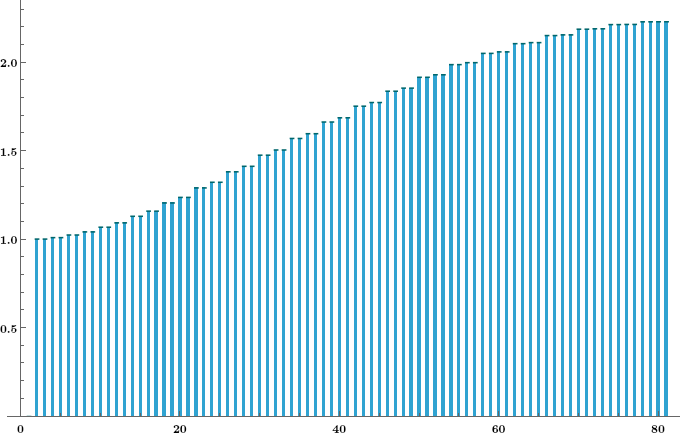} 
} \\
\resizebox{0.3\textwidth}{!}{%
  \includegraphics{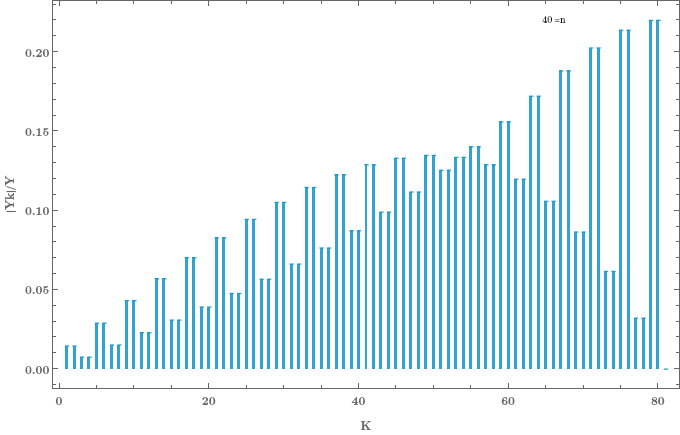}
}
     \caption{The upper plot shows the mass distribution and the lower plot shows the coupling strength produced between SM neutrino and $K^{th}$ mass mode for the NNN clockwork scenario with n = 40 clockwork fermions with $q = q' = -1$.} \label{NNN-mass}  
\end{figure}

The KK mass spectrum for Clockwork gears and their coupling strength produced between $K^{th}$ mass mode with SM neutrino before Higgs achieves vev is shown in Fig. \ref{NNN-mass}. The parameters chosen for this are n = 40 sites, with coupling parameters $q=-1$ and $q'=-1$. Fig. \ref{NNN_local} demonstrates the localization of different eigenvectors in CW fields on different sites for these parameters. The localisation of the right-handed 0-mode is clearly evident from the figure.
BP for the NNN-CW model that survives the current phenomenological constraints are listed in Table \ref{tab:bp_points_2}.

\begin{table}[h!]
\caption{BP points for the model with $q_i = q, q'_i = q' \hspace{0.5cm} \forall i \in \{1, 2, \ldots, n \}$ producing neutrino mass in agreement with current experimental data.}
\label{tab:bp_points_2} % Give a unique label
\centering
\begin{tabular}{lllll}  % Changed to five left-aligned columns
\hline\noalign{\smallskip}
\textbf{Hierarchy} & \textbf{n} & \textbf{$q$} & \textbf{$q'$} & \textbf{Yukawa Couplings} \\
\noalign{\smallskip}\hline\noalign{\smallskip}
Normal & 15 & $-5$ & $-5$ & $ \{y_1, y_2, y_3 \} = \{0.1, 0.101, 0.13 \}$ \\
\noalign{\smallskip}\hline\noalign{\smallskip}
Inverted & 15 & $-5$ & $-5$ & $ \{y_1, y_2, y_3 \} = \{0.1, 0.101, 0.06 \}$ \\
\noalign{\smallskip}\hline
\end{tabular}
\end{table}
\subsection{Completely Non-local CW}
\begin{figure}
   \centering
\resizebox{0.5\textwidth}{!}{%
  \includegraphics{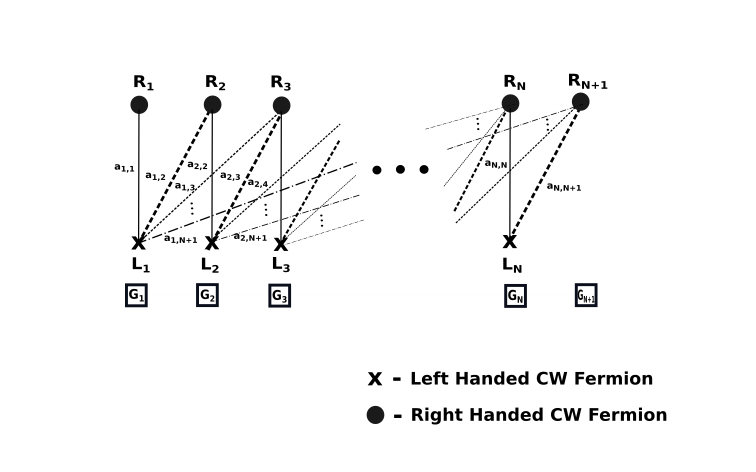} 
}    \caption{ CW with all neighbouring interactions. In CW notations, $a_{i,i}$ = $m_i$, $a_{i,i+1}$ = $m_iq_i^{(1)}$, $a_{i,i+2}$ = $m_iq_i^{(2)}$, $\hdots$, $a_{i,i+n}$ = $m_iq_i^{(n)}$. } \label{NnCW}
\end{figure}

In this further extension scenario, we will consider fully non-local theory spaces i.e., theory spaces where the matter fields of each group are connected via link fields to the matter fields of every other group. The underlying Hamiltonian is still considered as rectangular, implying that the number of left chiral fermions is not equal to the number of right chiral fermions. The CW nature of theory space is retained. The theory space is diagrammatically shown in Fig. \ref{NnCW}. Hamiltonian for this extension can be written as
 \begin{align}
    \mathcal{H}_{i,j} = \sum_{k=1}^{n+1} a_{i,k} \delta_{i,j-k+1} 
\end{align}
with i $\in$ $\{1,2,...n\}$ and j $\in$ $\{1,2,...n+1\}$. Using the CW notation to write the new physics Lagrangian \cite{ben2019generalized}, one gets
\begin{align} & \mathcal{L}_{NP} = \mathcal{L}_{kin} 
  - \sum_{i=1}^{n} m_i \overline{L_{i}}R_i - \sum_{i=1}^{n} m_iq_i^{(1)} \overline{L_{i}}R_{i+1} - \nonumber \\
  & \sum_{i=1}^{n-1} m_iq_i^{(2)} \overline{L_{i}}R_{i+2} - \sum_{i=1}^{n-2} m_iq_i^{(3)} \overline{L_{i}}R_{i+3} - \nonumber \\
  &  \sum_{i=1}^{n-3} m_iq_i^{(4)} \overline{L_{i}}R_{i+4} + \hspace{0.4cm} \hdots \hspace{0.4cm} - \sum_{i=1}^{n-(n-1)} m_iq_i^{(n)} \overline{L_{i}}R_{i+n}  + h.c. \nonumber \\
  &  = \mathcal{L}_{kin} 
  - \sum_{i=1}^{n} m_i \overline{L_{i}}R_i - \sum_{k=1}^n \sum_{i=1}^{n-k+1} m_iq_i^{(k)} \overline{L_{i}}R_{i+k}  + h.c. 
\end{align}
In this model, the mass matrix for right-handed CW fermions will also have a 0 mode as per the theorem since the Hamiltonian is rectangular. The matrix for this Hamiltonian in $\{\overline{L}_1,\overline{L}_2,...\overline{L}_n \} $
 and  $\{R_1,R_2,....R_{n+1} \}$ basis is given by 

 \begin{align*}
\scalebox{0.8}{$
M_{CW} = \begin{bmatrix}
    \makebox[1.5em][c]{$a_{1,1}$} & \makebox[1.5em][c]{$a_{1,2}$} & \makebox[1.5em][c]{$a_{1,3}$} & \makebox[1.5em][c]{$a_{1,4}$} & \dots & \makebox[1.5em][c]{$a_{1,n+1}$} \\
    \makebox[1.5em][c]{0} & \makebox[1.5em][c]{$a_{2,2}$} & \makebox[1.5em][c]{$a_{2,3}$} & \makebox[1.5em][c]{$a_{2,4}$} & \dots & \makebox[1.5em][c]{$a_{2,n+1}$} \\
    0 & 0 & \makebox[1.5em][c]{$a_{3,3}$} & \makebox[1.5em][c]{$a_{3,4}$} & \dots & \makebox[1.5em][c]{$a_{3,n+1}$} \\
    \vdots & \vdots & \vdots & \ddots & \ddots & \vdots \\
    \makebox[3em][c]{0} & \makebox[3em][c]{0} & \dots & \phantom{-}\makebox[3em][c]{$a_{n-1,n-1}$} & \phantom{-}\makebox[3em][c]{$a_{n-1,n}$} & \phantom{-}\makebox[3em][c]{$a_{n-1,n+1}$} \\
    0 & 0 & 0 & \dots & \phantom{-}\makebox[1.5em][c]{$a_{n,n}$} & \phantom{-}\makebox[1.5em][c]{$a_{n,n+1}$}
\end{bmatrix}_{n \times n+1}
$}
\end{align*}
In CW notations, $a_{i,i}$ = $m_i$, $a_{i,i+1}$ = $m_iq_i^{(1)}$, $a_{i,i+2}$ = $m_iq_i^{(2)}$, $\hdots$, $a_{i,i+n}$ = $m_iq_i^{(n)}$. The $K^{th}$ component for null space basis of CN-CW (completely non-local clockwork) in the limiting case, $a_{i,i+k}$ = $a_k$ $\forall$ i, is given by
\begin{align}
   \Lambda_0^K = \sum_{\{k_1,\hdots,k_n\}}^{}\frac{(k_1+k_2+\hdots +k_n)!}{k_1!\hdots k_n!}&\frac{(-a_n)^{k_n}\hdots (-a_1)^{k_1}}{a_0^{k_1+k_2+\hdots +k_n}}   \\
   with \hspace{2cm} nk_n \hdots + 2k_2 + k_1& =  n + 1 - K \\
   \Lambda_0 = \{\Lambda_0^1,\Lambda_0^2,\hdots,\Lambda_0^{n+1}\} \nonumber
\end{align}
and $k_1, k_2, \hdots ,k_n$ $\in$ $\mathbb{N_0} = \{0,1,2,3,4, ... \}.$ 
The normalized 0-mode $\Tilde{\Lambda}_0$ is given by $\mathcal{N}_0\Lambda_0$, with $\mathcal{N}_0$ representing the normalizing
factor. Using $\Tilde{\Lambda}_0^\dagger \Tilde{\Lambda}_0$ = 1, we get
$$ \mathcal{N}_0 = \frac{1}{\sqrt{\sum_{i=1}^{n+1}\Lambda_0^{i*} \Lambda_0^{i}}}$$
To compare with CW, it took n = 40 gears with a = 1, q = -2 to produce O(1) eV mass from the TeV scale, here it can be done with n = 25 for $a_0$ = 1, $a_1$ = -2, $a_2$ = -3 and $a_i$ = -2, i $\in$ [3,n]. However, compared to the NNN CW model, this model does not give orders of extra suppression unless hierarchy in the fundamental parameters $a_i$s are introduced for $|q| >$1. For the coupling strength with unity mod, the combinatorics factor for CN-CW will be much larger as compared to NNN-CW and there will again be larger suppression which will increase with n. Table \ref{tab:clockwork_model_comparison} compares CW, NNN-CW and CN-CW for unity coupling strength based on the amount of suppression produced.\\
Now, same as earlier, to study the effect of SM neutrino coupling with CW fermions perturbatively, we will consider the interaction term in Lagrangian of the same form as before eq.\eqref{lint}.
\begin{equation}
\mathcal{L}_{int}=-Y \widetilde{H} \bar{L}_{L} R_{n+1}+\text { h.c. }
\end{equation}
The Dirac mass matrix in the basis $\{\overline{\nu},\overline{L}_n, \overline{L}_{n-1},\hdots \overline{L}_1 \}$ and $\{R_{n+1}, R_n, R_{n-1},\hdots R_1 \}$ for Y = 0 is given by
\[ \Tilde{M} = m \begin{bmatrix}\makebox[1.5em][c]{0} & \makebox[1.5em][c]{0} & 0 & 0 & \dots & 0 \\
 q & 1 & 0 & 0 & \dots & 0 \\
q &  q & \makebox[1.5em][c]{1} & 0 & \dots & 0 \\
q & q & q & 1 & \hdots & 0 \\
\vdots & \vdots & \vdots & \ddots & \ddots & \vdots \\
q & q & q & \dots & q & \makebox[1.5em][c]{1}
\end{bmatrix}_{(n+1) \times (n+1)}
\]
 for simplicity, the case with the same value for all couplings is considered. The masses for right-handed CW fermions will be again given by $m\sqrt{\Tilde{\lambda_i}} $, with  $\Tilde{\lambda_i}$ denoting the eigenvalues of the matrix $\Tilde{M^\dagger}\Tilde{M}$/$m^2$.
{
\tiny
\[ \Tilde{M^\dagger}\Tilde{M} = m^2 \begin{bmatrix}
    \makebox[5.5em][c]{$nq^2$} & \makebox[6em][c]{$q+(n-1)q^2$} & \makebox[6em][c]{$q+(n-2)q^2$} &  \dots & \makebox[4em][c]{$q+q^2$} & \makebox[.5em][c]{$q$} \\ 
    \makebox[4em][c]{$q+(n-1)q^2$} & \makebox[3.5em][c]{$1+(n-1)q^2$} & \makebox[4em][c]{$q+(n-2)q^2$} &  \dots & \makebox[4em][c]{$q+q^2$} & \makebox[.5em][c]{$q$} \\
    \makebox[4em][c]{$q+(n-2)q^2$} & \makebox[1.5em][c]{$q+(n-2)q^2$} & \makebox[4.2em][c]{$1+(n-2)q^2$}  & \dots & \makebox[4em][c]{$q+q^2$} & \makebox[.5em][c]{$q$} \\
    \vdots & \vdots & \vdots &  \ddots & \vdots & \vdots\\
    \makebox[4em][c]{$q+q^2$} & \makebox[4em][c]{$q+q^2$} & \dots &  \makebox[1.5em][c]{$q+q^2$} & \makebox[3em][c]{$1+q^2$} & \makebox[.5em][c]{$q$} \\
    \makebox[4em][c]{$q$} & \makebox[4em][c]{$q$} & \dots & \makebox[1.5em][c]{$q$} & \makebox[3em][c]{$q$} & \makebox[.5em][c]{$1$}
\end{bmatrix}_{(n+1) \times (n+1)} \]
}
For non-zero coupling Y, the eigenstates and eigenvalues change depending on the value of Y. Consider p to be the perturbative coupling strength then the Dirac mass matrix $\Tilde{M}$ becomes
\[ M = m \begin{bmatrix}\makebox[1.5em][c]{p} & \makebox[1.5em][c]{0} & 0 & 0 & \dots & 0 \\
 \makebox[1.5em][c]{q} & \makebox[1.5em][c]{1} & 0 & 0 & \dots & 0 \\
q &  \makebox[1.5em][c]{q} & \makebox[1.5em][c]{1} & 0 & \dots & 0 \\
q & q & q & 1 & \hdots & 0 \\
\vdots & \vdots & \vdots & \ddots & \ddots & \vdots \\
q & q & q & \dots & \makebox[1.5em][c]{q} & \makebox[1.5em][c]{1}
\end{bmatrix}_{(n+1) \times (n+1)}
\]
The right-handed fermions have masses m$\sqrt{\lambda_i}$ with $\lambda_i$ being the eigenvalues of the following mass matrix:-
{\tiny
\[ \frac{M^\dagger M}{m^2} =  \begin{bmatrix}
  \makebox[6em][c]{$p^2+nq^2$} & \makebox[6em][c]{$q+(n-1)q^2$} & \makebox[6em][c]{$q+(n-2)q^2$} &  \dots & \makebox[4em][c]{$q+q^2$} & \makebox[1em][c]{$q$} \\ 
  \makebox[3em][c]{$q+(n-1)q^2$} & \makebox[3.5em][c]{$1+(n-1)q^2$} & \makebox[4em][c]{$q+(n-2)q^2$} &  \dots & \makebox[4em][c]{$q+q^2$} & \makebox[1em][c]{$q$} \\
  \makebox[4em][c]{$q+(n-2)q^2$} & \makebox[1.5em][c]{$q+(n-2)q^2$} & \makebox[4.2em][c]{$1+(n-2)q^2$} &  \dots & \makebox[4em][c]{$q+q^2$} & \makebox[1em][c]{$q$} \\
  \vdots & \vdots & \vdots &  \ddots & \vdots & \vdots \\
  \makebox[4em][c]{$q+q^2$} & \makebox[4em][c]{$q+q^2$} & \dots &  \makebox[1.5em][c]{$q+q^2$} & \makebox[3em][c]{$1+q^2$} & \makebox[1em][c]{$q$} \\
  \makebox[4em][c]{$q$} & \makebox[4em][c]{$q$} & \dots &  \makebox[1.5em][c]{$q$} & \makebox[3em][c]{$q$} & \makebox[1em][c]{$1$}
\end{bmatrix}_{(n+1) \times (n+1)} \]
}

Again as in previous cases, we can write this matrix as the sum of the Dirac matrix with 0 neutrino coupling and a perturbation matrix.
$$ M^\dagger M = \Tilde{M^\dagger}\Tilde{M} + \delta M^2 $$
with perturbation matrix given by
\[ \delta M^2 = m^2 \begin{bmatrix}
  p^2 & \boldsymbol{0}_{1 \times n} \\
  \boldsymbol{0}_{n \times 1} & \boldsymbol{0}_{n \times n}
\end{bmatrix}_{(n+1) \times (n+1)} \]

The perturbation matrix is independent of the CW coupling strength parameters the same as in earlier variants of CW. And it depends on the strength of the Standard Model (SM) field coupling. As a result, the leading-order corrections to the eigenvalues of the matrix are proportional to the SM field coupling strength parameter, p.
$$
\delta \lambda_i=\left\langle \Lambda^{(i)}\left|\frac{\delta M^2}{m^2}\right| \Lambda^{(i)}\right\rangle=p^2 f(q)=O(p^2)
$$
$\Lambda_i$ denotes eigenvectors of the unperturbed matrix for the CN-CW case. In a general scenario, $f(q^{(1)},q^{(2)},q^{(3)},...,q^{(n)})$ will be a function depending on $q^{(i)}$ coupling strength. Once again the leading order corrections are of the second order with respect to p.

\begin{figure}
\centering
\resizebox{0.3\textwidth}{!}{%
  \includegraphics{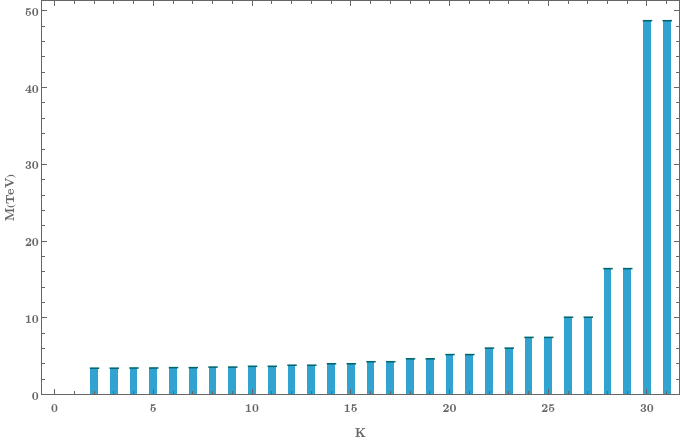} 
} \\
\resizebox{0.3\textwidth}{!}{%
  \includegraphics{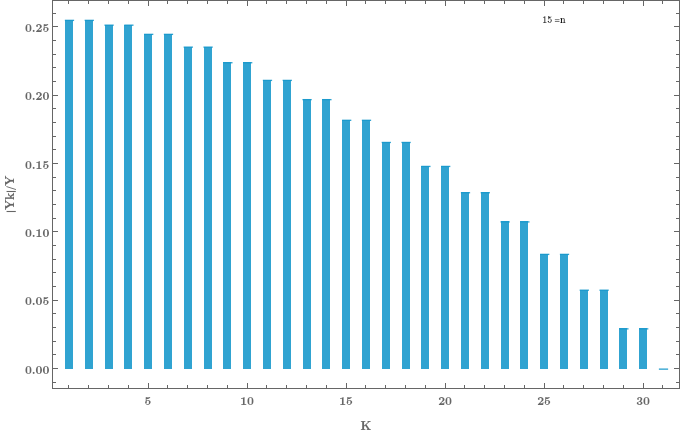}
}
     \caption{ The upper plot shows the mass distribution and the lower plot shows the coupling strength produced between $K^{th}$ mass mode and SM neutrino for the completely non-local clockwork scenario with BP parameters used are mentioned in table \ref{tab:bp_points_3}.} \label{CNCW-mass}
\end{figure}

\begin{figure}
\centering
\resizebox{0.3\textwidth}{!}{%
  \includegraphics{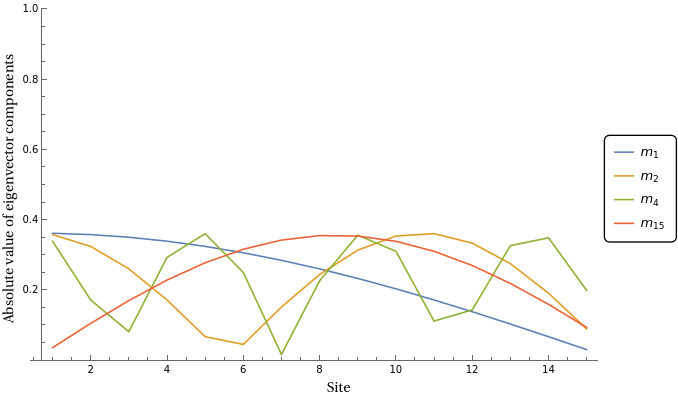} 
} \\
\resizebox{0.3\textwidth}{!}{%
  \includegraphics{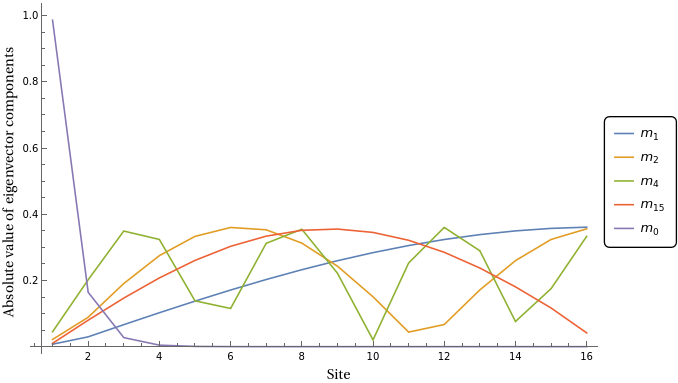}
}
    \caption{ The upper plot shows the absolute value of left-handed mass eigenvectors in terms of CW fields and the lower plot for right-handed mass eigenbasis before the Higgs achieves vev for n = 40 with $q = q' = -1$. The localisation of the 0-mode with these parameters is evident from the plot for $m_0$ mode. } \label{CNCW-local}
\end{figure}

The KK mass spectrum for Clockwork gears and their coupling strength produced between $K^{th}$ mass mode and SM neutrino for benchmark points is shown in Fig. \ref{CNCW-mass}. 
Fig. \ref{CNCW-local} demonstrates the localization of different eigenvectors in CW fields on different sites before Higgs achieves vev. The BP parameters for this model are listed in Table \ref{tab:bp_points_3}.

\begin{table}[h!]
\caption{BP points for the model with $q_i^j = q \hspace{0.2cm} \forall \hspace{0.2cm} i,j$ producing neutrino mass in agreement with experimental data.}
\label{tab:bp_points_3} % Give a unique label
\centering
\begin{tabular}{llll}  % Changed to four left-aligned columns
\hline\noalign{\smallskip}
\textbf{Hierarchy} & \textbf{n} & \textbf{$q$} & \textbf{Yukawa Couplings} \\
\noalign{\smallskip}\hline\noalign{\smallskip}
Normal & 15 & $-5$ & $ \{y_1, y_2, y_3 \} = \{0.1, 0.102, 0.155 \}$ \\
\noalign{\smallskip}\hline\noalign{\smallskip}
Inverted & 15 & $-5$ & $ \{y_1, y_2, y_3 \} = \{0.1386, 0.14, 0.08 \}$ \\
\noalign{\smallskip}\hline
\end{tabular}
\end{table}
As is evident from the BP of CN-CW in Table \ref{tab:bp_points_3}, the difference in hierarchy produced in this model is not significant from the NNN-CW model.\\
Apart from retaining the clockwork nature of left-right chiral fields in non-local extensions, we can consider both-sided non-local CW extensions too. Hamiltonian for this scenario is given by \cite{PhysRevD.103.015001}
\begin{align}
\left(\mathcal{H}_{\text {long-range }}\right)_{j, k}=a_{j} \delta_{j, k}+\frac{b}{r^{|j-k|}}\left(1-\delta_{j, k}\right), \label{27}
\end{align}
The Hamiltonian considered is long-range hopping strength decaying Hamiltonian. Dirac mass matrix for this Hamiltonian in $\{\overline{L}_1,\overline{L}_2,...\overline{L}_n \} $
and  $\{R_1,R_2,....R_{n+1} \}$ basis is given by
\[ M_{\text{long-range}} = 
\begin{bmatrix}
  \makebox[1.5em][c]{$a_1$} & \frac{b}{r} & \frac{b}{r^2} & \frac{b}{r^3} & \dots & \frac{b}{r^{n}} \\
  \frac{b}{r} & \makebox[1.5em][c]{$a_2$} & \frac{b}{r} & \frac{b}{r^2} & \dots & \frac{b}{r^{n-1}} \\
  \frac{b}{r^2} & \frac{b}{r} & \makebox[1.5em][c]{$a_3$} & \frac{b}{r} & \dots & \frac{b}{r^{n-2}} \\
  \vdots & \vdots & \vdots & \ddots & \ddots & \vdots \\
  \frac{b}{r^{n-2}} & \frac{b}{r^{n-1}} & \dots & \makebox[1.5em][c]{$a_{n-1}$} & \frac{b}{r} & \frac{b}{r^2} \\
  \frac{b}{r^{n-1}} & \frac{b}{r^{n-2}} & \frac{b}{r^{n-3}} & \dots & \makebox[1.5em][c]{$a_{n}$} & \frac{b}{r}
\end{bmatrix}_{n \times (n+1)}
\]

\begin{figure}[h]
    \centering
    \resizebox{0.4\textwidth}{!}{%
  \includegraphics{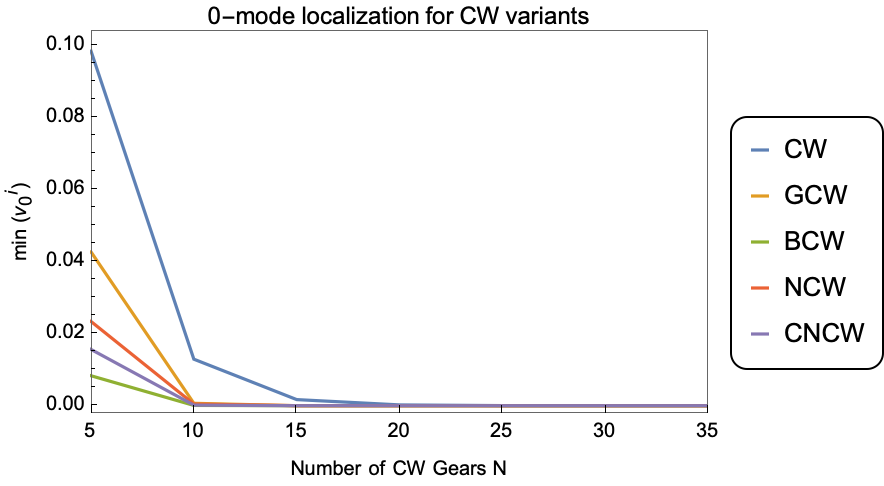} 
}
   \resizebox{0.4\textwidth}{!}{%
  \includegraphics{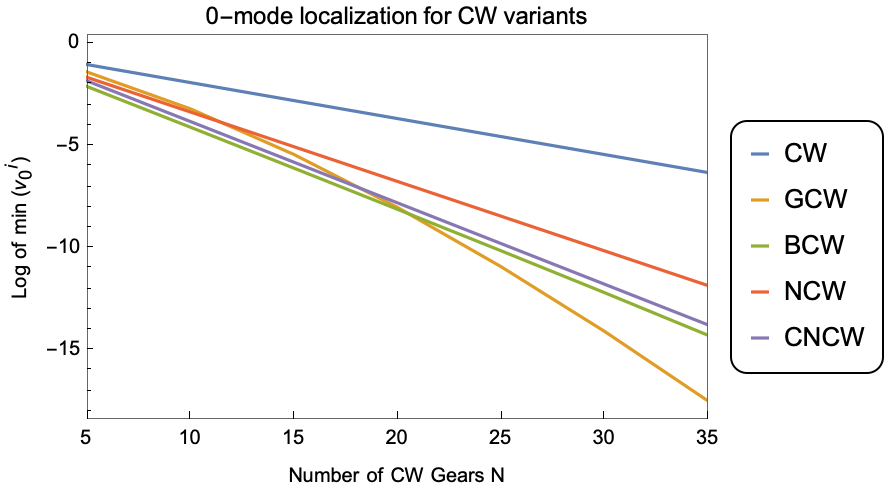} 
}
    \caption{For these plots, the scenario considered is q=-1.5 with q'=-1/4 for BCW and $q^i$=q for NCW and CNCW and $q_i=q-0.1 \times i$ for GCW. The X-axis represents the number of CW gears considered. } \label{0-mode comparison}
\end{figure}

For n = 2, in the limiting case, right-hand fermionic eigenvalues are
\begin{align}
\scalebox{0.95}{$
    \lambda_i =  \Bigl\{0,\frac{-b \sqrt{16 a^2 r^6+16 a b r^4+b^2 r^4+2 b^2 r^2+b^2}+2 a^2 r^4+3 b^2 r^2+b^2}{2 r^4}, \nonumber
$}
\end{align}
\begin{align}
\scalebox{1}{$
    \frac{b \sqrt{16 a^2 r^6+16 a b r^4+b^2 r^4+2 b^2 r^2+b^2}+2 a^2 r^4+3 b^2 r^2+b^2}{2 r^4}\Bigr\} \nonumber
$}
\end{align}
with 0-mode eigenvector given by
\begin{align}
    \Lambda_0 = \left\{-\frac{a b-b^2}{a^2 r^2-b^2},-\frac{a b r^2-b^2}{r \left(a^2 r^2-b^2\right)},1\right\}
\end{align}
Hence as b $\rightarrow$ $ar$, the suppression of 0-mode on the last site increases.
\begin{figure}
    \centering
    \resizebox{0.3\textwidth}{!}{%
  \includegraphics{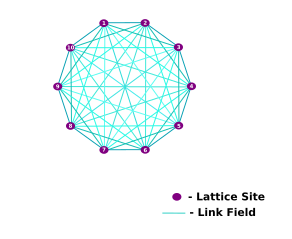} 
}
    \caption{ Completely Non-local interaction for n = 9 with both ways interaction. The distance between the two sites is not parallel to their coupling strength. }
\end{figure}
Now for n = 3, the 0-mode eigenvector has components given by
\begin{align}
   \Lambda_0 = & \Big\{-\frac{r \left(a^2 b-2 a b^2+b^3\right)}{a^3 r^4-2 a b^2 r^2-a b^2+2 b^3},-\frac{a b-b^2}{a^2 r^2+a b-2 b^2},\nonumber \\ & -\frac{a^2 b r^4-a b^2 r^2-a b^2-b^3 r^2+2 b^3}{r \left(a^3 r^4-2 a b^2 r^2-a b^2+2 b^3\right)},1 \Big\}
\end{align}
Under certain values of b, the suppression on a specific site will be sufficient to generate masses that are hierarchically smaller, in a natural manner.\\
Finally, we can compare the absolute value of the minimum component of the 0-mode produced by different variants of CW models. The result is shown in Fig. \ref{0-mode comparison} also the parameters considered for models are mentioned in Table \ref{tab:clockwork_models}. 
\begin{table}[h!]
\caption{Parameters and descriptions for different Clockwork models with varying number of CW gears.}
\label{tab:clockwork_models} % Give a unique label
\centering
\begin{tabular}{lll}  % Changed to three left-aligned columns
\hline\noalign{\smallskip}
\textbf{Model} & \textbf{Mass Parameters} & \textbf{Coupling Parameters} \\
\noalign{\smallskip}\hline\noalign{\smallskip}
CW  & $m_i = 1 $  &  $q_i = -1.5 $  \\
\noalign{\smallskip}\hline\noalign{\smallskip}
GCW & $m_i = 1 $  & $q_i = -1.5 - 0.1 \times i$  \\
\noalign{\smallskip}\hline\noalign{\smallskip}
BCW & $m_i = 1 $  & $q = -1.5$, $q' = -1/4$  \\
\noalign{\smallskip}\hline\noalign{\smallskip}
NNNCW & $m_i = 1 $  & $q^i = q = -1.5$   \\
\noalign{\smallskip}\hline\noalign{\smallskip}
CNCW & $m_i = 1 $  & $q^i = q = -1.5$   \\
\noalign{\smallskip}\hline
\end{tabular}
\end{table}
Even for higher values of q, the trend stays the same, i.e, variants of CW are producing bigger localization of 0-mode on a particular site compared to CW model and hence are more efficient to produce hierarchical masses. As shown in Fig. \ref{0-mode comparison}, the difference in localization strength increases with the number of gears introduced into the system. This is an expected result, as a greater number of gears leads to a larger combinatorial factor element in the 0-mode component.

Table \ref{tab:clockwork_model_comparison} demonstrates the presence of localisation of 0-mode phenomenon in non-local scenarios even for q = 1, while such localization is absent in the local ordinary clockwork (CW) case. As evident from the Table, the CN-CW model exhibits a steeper growth with increasing number of sites n compared to the NNN-CW model, due to the larger number of contributing combinations in the zero mode.  The explicit benchmark points (BPs) producing neutrino masses with 
$|q|=1$ for the NNN-CW and CN-CW scenarios are presented in Table \ref{tab:bp_points_4} and Table \ref{tab:bp_points_5}, respectively, in Appendix \ref{App:BP}.
\begin{table}[h!]
\caption{Comparison of different Clockwork model scenarios for various values of \( n \). The local and non-local coupling parameters are chosen as $a_{i,i}=1, a_{i,i+1}=q_i, a_{i,i+2}=q'_i, a_{i,i+3}=q''_i $ and so on. Parameter settings for each model are mentioned.}
\label{tab:clockwork_model_comparison}
\centering
\begin{tabular}{lccc}
\hline\noalign{\smallskip}
\textbf{\( n \)} & \textbf{CW} & \textbf{NNNCW} & \textbf{CNCW} \\
\noalign{\smallskip}
 & \( q_i = -1 \) & \( q_i = q'_i = -1 \) & \( q_i = q'_i = q''_i = \ldots = -1 \) \\
\noalign{\smallskip}\hline\noalign{\smallskip}
10  & $O(10^{-1})$ & $O(10^{-3})$  & $O(10^{-3})$ \\
20  & $O(10^{-1})$ & $O(10^{-5})$  & $O(10^{-6})$ \\
30  & $O(10^{-1})$ & $O(10^{-7})$  & $O(10^{-9})$ \\
40  & $O(10^{-1})$ & $O(10^{-9})$  & $O(10^{-12})$ \\
50  & $O(10^{-1})$ & $O(10^{-11})$ & $O(10^{-15})$ \\
60  & $O(10^{-1})$ & $O(10^{-13})$ & $O(10^{-18})$ \\
70  & $O(10^{-1})$ & $O(10^{-15})$ & $O(10^{-21})$ \\
\noalign{\smallskip}\hline
\end{tabular}
\end{table}

\section{Fine Cancellation In Dimensional Deconstruction Models}
Deconstruction models are the latticized extra spatial dimension models \cite{arkani2001constructing}. Since, these deconstruction models are the extra dimension models at low energy i.e, the physics produced by these two models at low energy converges, they are termed as dimensional deconstruction (DD) models. The DD models with moose diagrams having only nearest neighbour interactions for a finite number of groups are equivalent to the extra latticized spatial dimension picture for a finite number of sites at low energies. This low-energy extra-dimension physics is reproduced by considering structures among abstract groups at high energies. To set the notation we recap here DD which has been reviewed in several papers \cite{lane2003deconstructing}, \cite{hallgren2005phenomenological}.
For link fields $\Phi_{i,j}$, the action is given by \cite{arkani2001electroweak}
\begin{align}
\scalebox{0.8}{$
S_{link}=\int d^{4} x\left\{\sum_{i,j=1}^{N}\left[\operatorname{Tr}\left(\left(D_{\mu} \Phi_{i, j}\right)^{\dagger} D^{\mu} \Phi_{i, j}\right)-\frac{1}{4} F_{i, \mu \nu, a} F^{i, \mu \nu, a}\right]-V(\Phi)\right\} 
$}
\end{align}
where a = 1,2,...,$m^2$-1. The fields are considered to propagate in 4-dimensional spacetime. For SSB to take place, the potential that will produce non-zero \textit{vev} for link fields can be written as \cite{hallgren2005phenomenological}
\begin{equation}
\begin{aligned}
V(\Phi) &=\sum_{j=1}^{N}\left[-M''^{2} \operatorname{Tr}\left(\Phi_{i, j}^{\dagger} \Phi_{i, j}\right)+\lambda_{1} \operatorname{Tr}\left(\Phi_{i, j}^{\dagger} \Phi_{i, j}\right)^{2}\right.\\
&\left.+\lambda_{2}\left(\operatorname{Tr}\left(\Phi_{i, j}^{\dagger} \Phi_{i, j}\right)\right)^{2}+M^{\prime}\left(e^{\mathrm{i} \theta} \operatorname{det}\left(\Phi_{i, j}\right)+\text { h.c. }\right)\right]
\end{aligned}
\end{equation}
For $\lambda_1$, $\lambda_2$ and $M'$ $>$ 0 and $M''^2$ $<$ 0, this potential produces a Mexican hat shape and leads to non-zero \textit{vev} for the link field.
\begin{align}
  S_{matter} & = \sum_{i,j=1}^{N} \int d^{4} x\Bigl\{\bar{\psi}\left(i \gamma^{\mu} D_{\mu} \right) \psi + \Big(\overline{L_i}\Phi_{i, j} R_{j}+ \nonumber \\ & \overline{L_{j}} \Phi_{j, i} R_{i}\Big) 
   + \overline{L_j}MR_j  + h.c. \Bigr\} 
\end{align}
Depending on the \textit{vevs} of link fields various kinds of Hamiltonians and hence theory spaces are produced. If the \textit{vevs} of link fields connecting only the consecutive groups' matter fields are non-zero i.e, $<\Phi_{i,j}> $ $\neq 0$ only for j = i -1 and i+1 then one gets a local theory space if it is non-zero for other i and j than one gets non-local theory spaces. The local DD model is equivalent to the ADD model at low energies \cite{hallgren2005phenomenological}, which forms the underlying framework for implementing the fine-cancellation mechanism in this work.

\subsection{One Flavour Scenario}
The Hamiltonian in this scenario is the same as the uniform tight-binding model Hamiltonian \cite{anderson1958absence}. The mode localization is not present in this Hamiltonian so the wavefunctions are spread throughout the sites.\cite{hallgren2005phenomenological}
\begin{align}
    \mathcal{L}_{NP} = \mathcal{L}_{kin}  - \sum_{i,j=1}^{n} \overline{L_{i}}\mathcal{H}_{i,j}R_j  + h.c. 
\end{align}
with
\begin{align}
    \mathcal{H}_{i,j} =&\epsilon \delta_{i,j} - t(\delta_{i+1,j} + \delta_{i,j+1} )
\end{align}
$\epsilon$ $ \& $ $t$ are the new parameters of the model. This Hamiltonian produces several delocalized modes with eigenmasses and eigenvectors given by \eqref{6} $\&$ \eqref{7} 
\begin{align}
    \lambda_{k}=\epsilon -2t\cos \frac{k \pi}{n+1}, \label{6}
\end{align}
for $k \in\{1,2, \ldots, n\}$, and the corresponding $\chi_{j}^{(k)}$, eigenvectors are given by
\begin{align}
\chi_{j}^{(k)}=\rho^k \sin \frac{k j \pi}{n+1}, \hspace{1cm} j \in \{1,2, \ldots, n\}  \label{7}
\end{align}
where $\rho^k$ is the normalization factor for $k^{th}$ eigenvector.

There is no localization of any kind in this model. Since the Dirac mass matrix is symmetric, the rotation of the left and right modes will be identical as shown in Appendix\ref{app:linear_algebra}. The unitary transformation in diagonalisation will make sure different modes are orthogonal to each other. Since the rotation matrices are unitary, the inner product of different modes will be 0 i.e,
$$ \sum_{i=1}^nv_j^{i}v_k^{i} = \delta_{j,k} $$
For the SM neutrino interaction to BSM fields with SM Higgs consider the following interaction term in the Lagrangian \cite{arkani2001neutrino}:- \footnote{This interaction is the same as is considered in \cite{craig2018exponential}. This is similar to localizing left chiral and right chiral neutrinos on opposite branes in RS/ADD, depending on the Higgs profile.}
$$ \mathcal{L}_{int.}=  Y\Bar{\nu}_LHR_1 + Y\Bar{\nu}_RHL_n + h.c.  $$
The smallest mass mode for the final mass matrix with weaker couplings Y is given by
$$ m_0 \approx v^2 \sum_{i=1}^n \frac{v_1^i v_n^i}{\lambda_i} $$
with $v$ as the expectation value of the Higgs field and $v_n^i$ being the $i^{th}$ component of $n^{th}$ eigenvector. If $\lambda_i$ = $\lambda$ $\forall$ i, then $m_0$ $\rightarrow$ 0 from unitarity condition. 
For, 
\begin{align}
    \lambda_{i}&=\epsilon- 2t \cos \frac{i \pi}{n+1}, \nonumber \\ 
    \frac{\lambda_{i}}{\epsilon}&=1- \frac{2t}{\epsilon} \cos \frac{i \pi}{n+1}, 
\end{align}
hence for $t$ $\ll$ $\epsilon$, $\frac{\lambda_i}{\epsilon}$ $\rightarrow$ 1 $\forall$ i.  For this case, the approximate value of $m_0$ is given as
\begin{align}
m_0 &\approx v^2 \sum_{i=1}^n \frac{v_1^i v_n^i}{\lambda_i} \nonumber \\ 
 &= v^2 \sum_{i=1}^n \frac{v_1^i v_n^i}{\epsilon}\frac{\epsilon}{\lambda_i} \nonumber \\ 
 &= v^2 \frac{1}{\epsilon}\sum_{i=1}^n v_1^i v_n^i(1-x_i)^{-1} \nonumber \\ 
 &= \frac{v^2}{\epsilon}\sum_{i=1}^n v_1^i v_n^i + \frac{v^2}{\epsilon}\sum_{i=1}^n v_1^i v_n^i x_i + ... \nonumber \\ 
 m_0& \approx  \frac{v^2}{\epsilon}\sum_{i=1}^n v_1^i v_n^i x_i  
\end{align}
where, 
$$ x_i =  \frac{2t}{\epsilon} \cos \frac{i \pi}{n+1} $$
so $x_i$ $\rightarrow$ 0 $\Rightarrow$ $m_0$ $\rightarrow$ 0. This mechanism will work better for matrices whose eigenvalue spectrum has the same range order as in ADD models. For hierarchical spectra as in warped models, this will not work efficiently. As the separation between the magnitude of minimum and maximum eigenvalue increases, this mechanism's effectiveness in producing hierarchical scale decreases.

\begin{figure}
    \centering
    \resizebox{0.4\textwidth}{!}{%
  \includegraphics{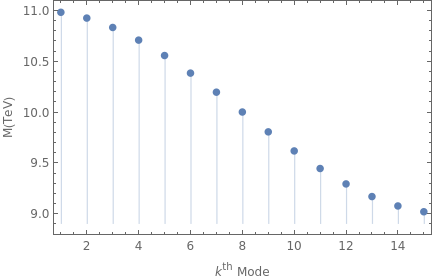} 
}
   \resizebox{0.4\textwidth}{!}{%
  \includegraphics{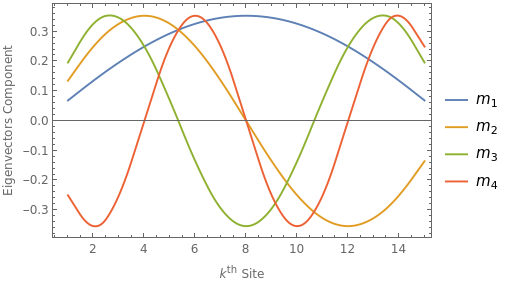} 
}
    \caption{ Figure shows the mass spectrum of modes (upper) and eigenvectors for the first four heavy modes (lower) for n = 15, $t = 0.5$ and $\epsilon = 10$.} \label{decon-mass}
\end{figure}

\begin{figure}
    \centering
    \resizebox{0.5\textwidth}{!}{%
  \includegraphics{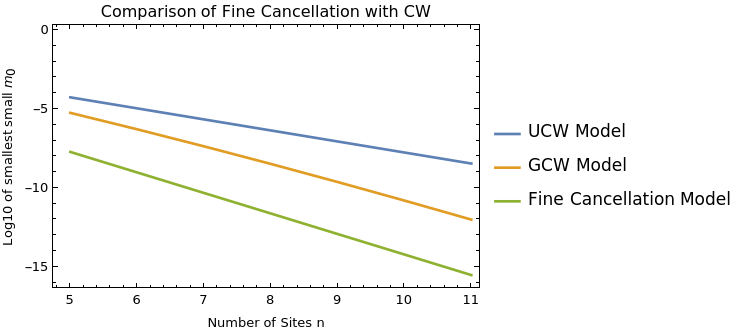} 
}
\caption{ Figure shows the comparison of the smallest mass-produced by UCW (uniform clockwork) eq.\eqref{GCW} with parameters $m_i$ = 1 and $q_i$ = -5 $\forall i \in \{1, n\}$, GCW (generalized clockwork) eq.\eqref{GCW} with parameters $m_i$ = 1 and $q_i=-5 - i$, $1 \leq i \leq n$ and Fine Cancellation model with $\epsilon$ = 10 and t = 0.5 for the varying number of sites n.} \label{comp-decon-cw}
\end{figure}

\begin{figure}
    \centering
    \resizebox{0.4\textwidth}{!}{%
  \includegraphics{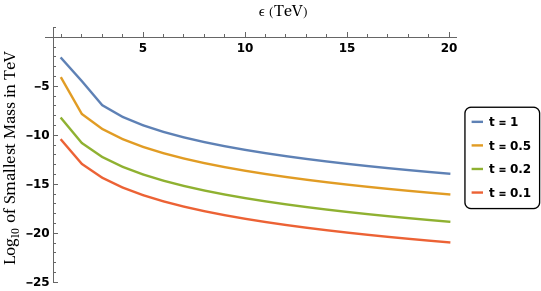} 
}
   \resizebox{0.4\textwidth}{!}{%
  \includegraphics{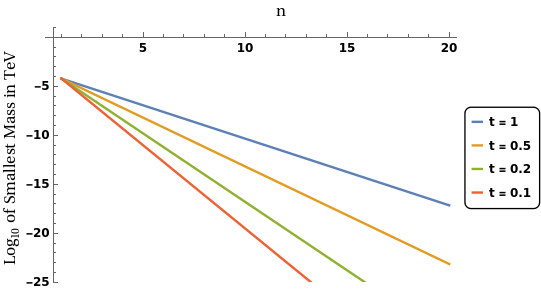} 
}
    \caption{ Figure shows the variation of log of smallest mass scale produced with the model scale parameter $\epsilon$ with n = 8 (upper) and with the number of sites/fields for each flavour (lower) with $\epsilon$ = 5.} \label{decon-ep-t}
\end{figure}

We find for $\epsilon$ = 10, t = 0.5 and n = 8, we get a small mass of the order 0.1 eV from the TeV scale i.e, $O(0^{12})$ magnitude smaller scale than the fundamental parameter scales of the theory. Fig. \ref{decon-mass} shows the mass spectra and eigenvectors for some massive modes for n = 15, $t = 0.5$ and $\epsilon = 10$. The figure demonstrates that modes are not localized and mass spectra are close to degenerate. Fig. \ref{comp-decon-cw} shows a comparison for the smallest mass scale produced between the DD model, uniform clockwork (UCW) and generalized clockwork (GCW) models. This figure shows that for the chosen parameters, as mentioned in Table \ref{tab:models_comparison}, the mass scales produced by DD is a few orders of magnitude smaller than both UCW and GCW for various values of sites. Though this is not true throughout the entire parameter space, there exist values of 
$\epsilon$ and t, where the fine-cancellation performs worse than in the (G)CW models. As expected, this performance depends on the degeneracy of the mass spectrum: the greater the degeneracy, the stronger the resulting hierarchy. Hence, the larger the $\epsilon/t$ ratio value, the finer the cancellation and hierarchy produced. Fig. \ref{decon-ep-t} shows the smallest mass scale produced by this model for varying $\epsilon$ and $t$ (left) and for varying sites n and $t$ (right). The figure shows that large values of n $\&$ $\epsilon$ and/or small values of $t$ produce a smaller mass scale, which is in agreement with the above understanding of the model.
\begin{table}[h!]
\caption{Parameters for comparison of UCW, GCW, and Fine Cancellation models in Fig. \ref{comp-decon-cw} with a varying number of sites \(n\).}
\label{tab:models_comparison} % Give a unique label
\centering
\begin{tabular}{lll}  % Changed to three left-aligned columns
\hline\noalign{\smallskip}
\textbf{Model} & \textbf{Mass} & \textbf{Coupling Parameters} \\
\noalign{\smallskip}\hline\noalign{\smallskip}
UCW & $m_i = 1$,  & $q_i = -5 $ \\
\noalign{\smallskip}\hline\noalign{\smallskip}
GCW & $m_i = 1$,  & $q_i = -5 - i $ \\
\noalign{\smallskip}\hline\noalign{\smallskip}
Fine Cancellation & $\epsilon = 10$,  & $t = 0.5$ \\
\noalign{\smallskip}\hline
\end{tabular}
\end{table}

\subsection{Three Flavour Scenario}
This can be extended to 3 flavour cases to account for all three SM active neutrino masses. The number of sites for each flavour is taken to be the same with differing flavour neutrinos coupling to the BSM fields. These varying couplings will produce different masses for different active neutrinos. The full 3 flavour Lagrangian for this scenario is given by
\begin{align}
    \mathcal{L}_{NP} =& L_{kin} -\sum_{i,j=1}^{n} \overline{L_{i}^{\alpha}}\mathcal{H}_{i,j}^{\alpha,\beta}R_j^{\beta} + h.c. \label{12} \\
     \mathcal{H}_{i,j}^{\alpha,\beta} =& \epsilon_i^{\alpha,\beta}\delta_{i,j} + t^{\alpha,\beta}(\delta_{i+1,j} + \delta_{i,j+1}) 
 \end{align}
with the interaction between different flavours of SM and BSM neutrino fields given by
\begin{align}
     \mathcal{L}_{Int.} = Y_1^{\alpha,\beta}\Bar{\nu}_L^{\alpha}HR_1^{\beta}  + Y_2^{\alpha,\beta}\Bar{\nu}_R^{\alpha}HL_n^{\beta}& + h.c. \nonumber
\end{align}
where $\alpha$ and $\beta$ are flavor index. For non-diagonal flavour Hamiltonian $H^{\alpha, \beta}$ and/or non-diagonal flavour Yukawa coupling $Y_1^{\alpha,\beta}$, $Y_2^{\alpha,\beta}$, this Lagrangian will produce mixing among three neutrino flavours. Here we are considering the case of diagonal flavour matrices hence no mixing is produced.
Table \ref{tab:bp_points_hierarchy} lists the benchmark points for the model that are consistent with neutrino mass constraints and remain viable under current experimental bounds.
\begin{table}[h!]
\caption{BP points for the model with $Y_1 = Y_2 = \text{diag}\{Y^{1,1}, Y^{2,2}, Y^{3,3} \}$ producing neutrino mass in agreement with experimental data.}
\label{tab:bp_points_hierarchy} % Give a unique label
\centering
\begin{tabular}{lllll}  % Changed to five left-aligned columns
\hline\noalign{\smallskip}
\textbf{Hierarchy} & n & \textbf{$\epsilon$} & \textbf{$t$} & \textbf{Yukawa Couplings} \\
\noalign{\smallskip}\hline\noalign{\smallskip}
Normal & 8 & 15 & $0.5$ &  { \tiny $ \{Y^{1,1}, Y^{2,2}, Y^{3,3} \} = \{0.1, 0.31, 0.7 \}$ } \\
\noalign{\smallskip}\hline\noalign{\smallskip}
Inverted & 8 & 15 & $0.5$ & { \tiny $ \{Y^{1,1}, Y^{2,2}, Y^{3,3} \} = \{0.7, 0.706, 0.1 \}$} \\
\noalign{\smallskip}\hline
\end{tabular}
\end{table}

\section{Phenomenological Signatures}
The coupling introduced between BSM fields and SM fields to explain the neutrino mass production will have contributions to SM processes and hence can be phenomenologically tested.
\begin{equation}
\mathcal{L}_{int}=-Y \widetilde{H} \bar{L}_{L} R_{n+1}+\text { h.c. }
\end{equation}
After changing the basis to the mass eigenvectors $\chi_k$ of the matrix $M_{CW}$, the Lagrangian interaction term can be rewritten as \cite{ibarra2018clockwork}
\begin{equation}
\mathcal{L}_{\text {int }}=-Y \bar{L}_{L} \widetilde{H} \mathcal{U}_{n+1, k} \chi_{k} \equiv-\sum_{k=1}^{n+1} Y_{k} \bar{L}_{L} \widetilde{H} \chi_{k} \label{43}
\end{equation}
One of the biggest observable signatures this extra vertex introduces is the charged lepton flavour violation branching ratio. In SM, the contribution to $BR(\mu \rightarrow e \gamma)$ comes via higher-order loops and hence is very tiny $O(10^{-55})$ but these new vertices in the Lagrangian lead to drastic increment to this value of BR and hence puts one of the most stringent bound and also is one of the best channels to discover new physics. These new BSM fields can directly manifest themselves in collider experiments as the missing energies or displaced vertex as in LPL depending on their decay widths. These BSM neutrinos will also affect the cross-section of various processes at LEP and Hadron colliders via new Feynman diagrams as explained in the below section. Since, the gears-SM interaction term eq.(\ref{43}) couples Higgs with these SM-BSM fields, this leads to gears contributing to Higgs self-energy and also to possible Higgs decay width depending on the masses of new fields as discussed below. The other possible phenomenological signatures are discussed in \cite{giudice2018clockwork}.
\subsection{Branching Ratio of FCNC}
\subsubsection{For Clockwork Variant Models}
\begin{figure}
    \centering
    \resizebox{0.3\textwidth}{!}{%
  \includegraphics{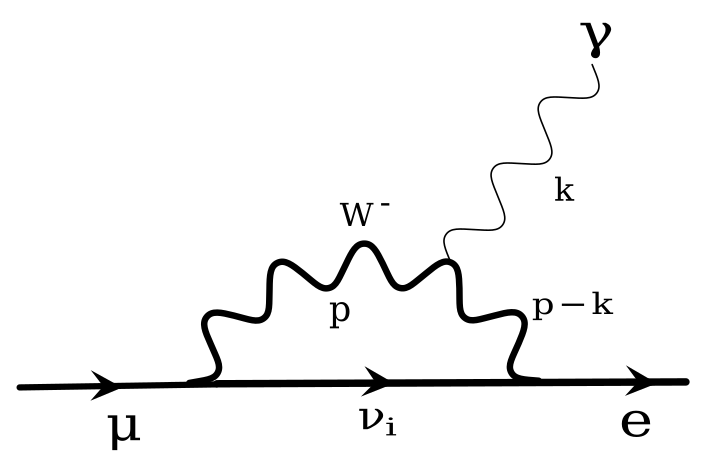} 
}
    \caption{ Feynman diagram depicting the lepton flavour change channel. } \label{BR}
\end{figure}

The Feynman diagram contributing to FCNC $BR(\mu \rightarrow e \gamma)$ is shown in Fig. \ref{BR}. Since these models introduce extra heavy neutral leptons as the propagator in this diagram, the BR will deviate significantly from the SM FCNC BR.
In the clockwork models, the $\mu \rightarrow e \gamma$ branching ratio is 
\cite{PhysRevLett.45.1908},\cite{ma1981exact}
\begin{align}
\operatorname{Br}(\mu \rightarrow e \gamma) & =\frac{3 \alpha}{8 \pi}|\mathcal{A}|^2, \label{44} \\
\mathcal{A} & =\sum_{\alpha=1}^3 \sum_{j=1}^{N+1} V_{\mu \alpha} V_{e \alpha}^*\left|\left(U_{L \alpha}\right)^{0 j}\right|^2 F\left(\frac{m_{j, \alpha}^2}{m_W^2}\right), \label{45}
\end{align}

\begin{figure}[h]
    \centering
    \resizebox{0.4\textwidth}{!}{%
  \includegraphics{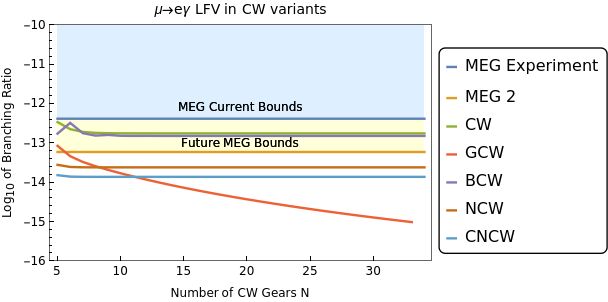} 
}
   \resizebox{0.4\textwidth}{!}{%
  \includegraphics{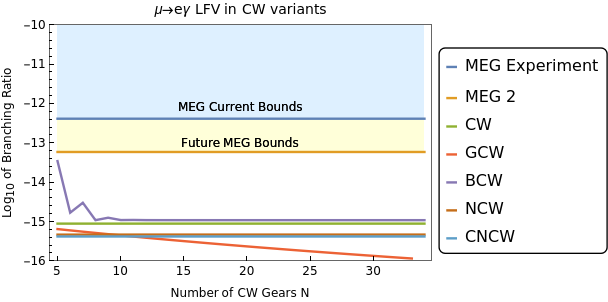} 
}
    \caption{ For these plots, the scenario considered is p=0.01, q=-1.7 with q'=q/2.5 for BCW (upper) and q=-5 with q'=q/3 for BCW (lower) and $q^i$=q for NCW(or NNNCW) and CNCW and $q_i=q-0.1 \times i$ for GCW. The number of CW gears considered is on the x-axis. } \label{comparison-cw}
\end{figure}

with
\begin{align}
\scalebox{0.95}{$
F(x)=\frac{1}{6(1-x)^4}\left(10-43 x+78 x^2-49 x^3+4 x^4+18 x^3 \log x\right)
$}
\end{align}
Fig. \ref{comparison-cw} compares BR obtained for Lepton flavour violation in the $\mu \rightarrow e \gamma$ process with various CW scenarios.

For bigger values of hopping strengths, the BR decreases for all CW variants and most of these variants stabilise after a certain value of the number of CW gears considered. Hence energy scale at the order of 10 TeV with this set of parameters will survive the experimental constraints put up by the MEG experiment $BR(\mu \rightarrow e \gamma)$ $\leq$ $4.2 \times$ $10^{-13}$ \cite{bao2016search} and are within reach of upcoming experiments for some parameter space with sensitivity up to $6 \times$ $10^{-14}$ \cite{MEGII:2021fah}. 
\subsubsection{For Fine-Cancellation Model} 
\begin{figure}
    \centering
    \resizebox{0.4\textwidth}{!}{%
  \includegraphics{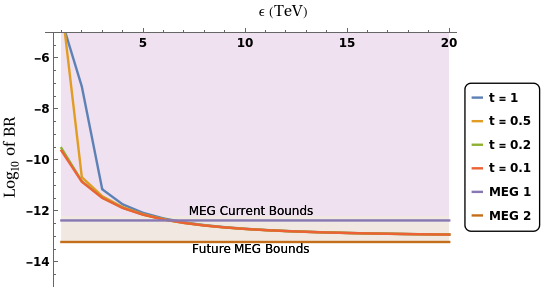} 
}
    \caption{ This plot shows the log of BR with base 10 vs model scale parameter $\epsilon$ with n = 8 BP and various hopping couplings in the fine-cancellation model. } \label{fine-BR}
\end{figure}

Similar to the clockwork model, in this model the FCNC branching ratio for $\mu \rightarrow e \gamma$ is given by the above equations \eqref{44}, \eqref{45} since the mixing in the considered scenario is not produced by the model. In the case of mixing being also produced by this model, the BR expression would have been given by a similar expression \cite{cheng1994gauge}. The elements considered for the mixing matrix were experimental PMNS values with unitarity conditions to good significant digits. The variation of BR for the benchmark point (BP) with varying numbers of sites is shown in the Fig. \ref{fine-BR}. 
As can be seen from the plot, the BP survives the current MEG bounds but they are within the reach of future MEG constraints. The pink region denotes the ruled-out parameter space region.

\subsection{Collider Signatures}
The new BSM heavy neutral leptons will have impact on collider physics and their effects can be observed at a high enough energy collider as is elaborated in \cite{hong2019clockwork}. These new contributions are emerging from the weak sector. These extra contributions in weak currents from the massive BSM neutrinos will affect the cross sections observed at colliders. The SM charged current Lagrangian is
$$
\mathcal{L}_{\mathrm{CC}}=\frac{g}{a} W_\mu^{-} J_W^{\mu+}+\text { h.c., }
$$
where $g$ is the SM weak coupling, and
\begin{align}
J_W^{\mu+}=& \frac{1}{\sqrt{2}} \bar{e}_\alpha \gamma^\mu \nu_{L \alpha} \\
=& \sum_{j=1}^{n+1} \frac{\left(U_{L \alpha}\right)^{j}}{\sqrt{2}} \bar{e}_\alpha \gamma^\mu P_L \nu_{j,\alpha} .
\end{align}
with 
$$
\nu_{L \alpha} = \sum_{j=1}^{n+1} \left(U_{L \alpha}\right)^{j} P_L \nu_{j,\alpha}
$$
Here $\alpha$ represents flavour index, and $P_L=\frac{1-\gamma_5}{2}$ is the left-handed projection operator and $\left(U_{L \alpha}\right)^{j}$ is the component of $\alpha$ flavour neutrino on $j^{th}$ massive neutral lepton. Similarly, the neutral current will also have contributions from these massive neutrinos. In \cite{hong2019clockwork} authors have studied the effects of neutrinos in both these CC and NC channels at hadron ($p p \rightarrow 3 l +  \slashed{E}_T$) and lepton colliders ($e^+ e^- \rightarrow l \nu j j$) for various CW neutrinos masses and have given the signal-background distributions. The variants of CW models studied in this paper will have similar results as the fundamental mechanism is the same in all variants. The deviations will occur due to slight differences in the masses spectrum and coupling strengths produced in each variant.

\subsection{Higgs Decay Width  \&  Radiative Corrections}
The nonzero coupling of BSM particles such as clockwork fermions with SM Higgs field will inevitably lead to corrections in Higgs mass. The leading order corrections occur at the 1-loop level. Consider the uniform CW model 
\begin{align} \mathcal{L}_{CW} = \mathcal{L}_{kin} - \sum_{i}^{n-1} m \overline{L_{i}}R_i - \sum_{i}^{n-1} mq \overline{L_{i}}R_{i+1}  + h.c.
\end{align}
with SM interaction part of Lagrangian in CW basis as
\begin{equation}
\mathcal{L}_{int}=-Y \widetilde{H} \bar{L}_{L} R_{n}+\text { h.c. }
\end{equation}
This term is written in CW mass basis once the link fields achieve \textit{vev} as \eqref{43}. Alternatively one can use the basis { \small $N_L$ = $(\nu_L,N_{L1},N_{L2},...,,N_{Ln})$ } and {\small$N_R$ = $(N_{R0},N_{R1},N_{R2},...,,N_{Rn})$} with transformations
$$
\begin{aligned}
& N_{R k}=\frac{1}{\sqrt{2}}\left(\chi_k+\chi_{k+n}\right), \quad k=0, \ldots n \\
& N_{L k}=\frac{1}{\sqrt{2}}\left(-\chi_k+\chi_{k+n}\right), \quad k=1, \ldots n .
\end{aligned}
$$
to write it as \cite{ibarra2018clockwork}
\begin{equation}
\mathcal{L}_{int}=-\sum_{k=0}^{ n} Y_{k} \bar{L}_{L} \widetilde{H} N_{Rk} + h.c.
\end{equation}
with $\chi_i$ fields representing the mass eigenstates. Once the Higgs achieves a \textit{vev} and breaks the SM symmetry, the physical basis once again rotates. Using SVD the rotation of left $N_L$ = U$N'_L$ and right basis $N_R$ = V$N'_R$ is determined. The terms giving mass to $N_L$ and $N_R$ after Higgs achieves \textit{vev} are
\begin{align}
\mathcal{L}_{mass}&=-\overline{N_L} m_{\nu}^D N_R -\frac{v}{\sqrt{2}} \sum_{k=0}^{ n} Y_{k} \bar{\nu}_{L} N_{Rk} + h.c. \nonumber \\ & 
= -\overline{N_L} m_{\nu}^D N_R - \overline{N_L} M_{int} N_R
\end{align}
$m_{\nu}^D$ has the form of diagonal matrix and $M_{int}$ matrix has only one non-zero row in these basis. Using the fact that diagonalization of the matrix $m_{\nu}^D$ + $M_{int}$ with U and V unitary matrices does not guarantee the diagonalization of the individual matrix $m_{\nu}^D$ and $M_{int}$ by these same unitary matrices, we find that SM Higgs gets mass corrections also from the fermionic-loops in which running fermions are not the same. The interaction of Higgs with new fields is covered by the following lagrangian :
\begin{align}
\mathcal{L}_{h}&= -\frac{h}{\sqrt{2}} \sum_{k=0}^{ n} Y_{k} \bar{\nu}_{L} N_{Rk} + h.c. \nonumber \\ & 
=  - \frac{h}{v}\overline{N_L} M_{int} N_R + h.c.\nonumber \\ & 
=  - \frac{h}{v}\overline{N'_L} U^{\dagger}M_{int}V N'_R + h.c.
\end{align}
since $U^{\dagger}M_{int}V$ $\neq$ $M_D$ is not necessarily a diagonal matrix, if the coupling between the Higgs field and different fermions is strong enough it will have some implications on the Higgs hierarchy problem.
The Feynman diagrams in Fig. \ref{Feynman-loop} give the radiative corrections to Higgs mass at the 1-loop level:
\begin{figure}
    \centering
    \resizebox{0.4\textwidth}{!}{%
  \includegraphics{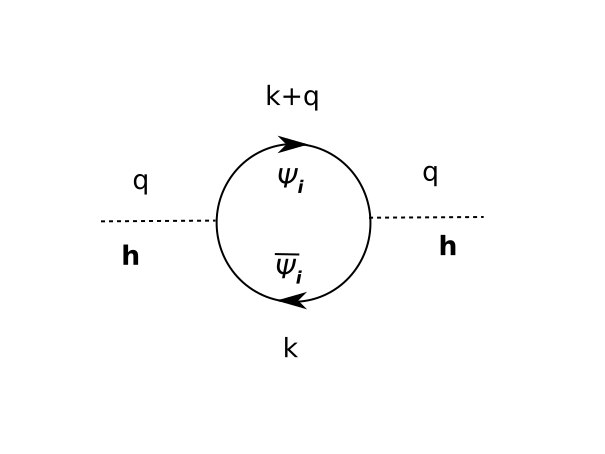} 
}
   \resizebox{0.4\textwidth}{!}{%
  \includegraphics{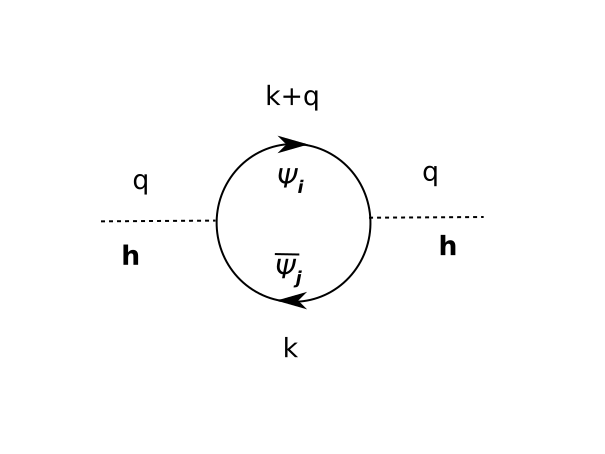} 
}
    \caption{ These Feynman diagrams show the 1-loop contribution of fermions in Higgs mass radiative corrections. The upper diagram shows it for the same fermions in the loop with $y_{ii}$ coupling, and the lower diagram shows it for different fermions in the loop with $y_{ij}$ coupling strength. } \label{Feynman-loop}
\end{figure}
The amplitude for 1-loop with the same fermions in the loop, $m_i$ = $m_j$ = m and $y_{ij}$ = $y_{ii}$ is given by
\begin{align}
\Pi_R(q, \mu)= N_f\frac{12 y_{ii}^2}{(4 \pi 
)^2}\left(\frac{q^2}{18}-\frac{m^2}{3}+\int_0^1 d x \Delta^2 \ln \left(\frac{\Delta^2}{\mu^2}\right)\right)
\end{align}
$N_f$ is the number of flavours. The contribution to Higgs's mass is proportional to coupling$\times$mass of fermions.
\begin{align}
    \delta m_h^2 \propto (y^2_{ii})(m^2)
\end{align}
Hence, it slightly worsens the Higgs boson mass hierarchy, similar to other TeV-scale BSM models. A similar result is obtained for different fermions in the loop. Calculation for this case is done in the Appendix \ref{app:one-loop}.

\begin{figure}
    \centering
    \resizebox{0.4\textwidth}{!}{%
  \includegraphics{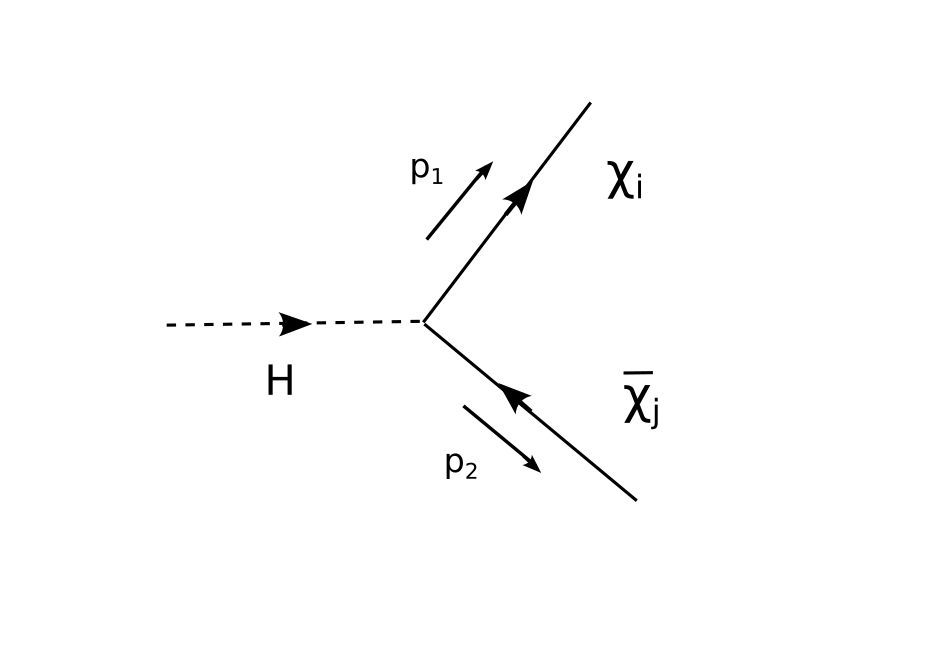} 
}
    \caption{ Feynman Diagram for Higgs decay.} \label{Higgs_decay}
\end{figure}

Finally, consider the impact on the Higgs decay width. Consider Higgs decaying to two particles $\chi_i$ and $\chi_j$ of masses $m_1$ and $m_2$ and four-momenta $p_1$ and $p_2$ with $M$ mass of Higgs and $q$ four-momentum. $H(q) \rightarrow \chi_i\left(p_1\right)+\chi_j\left(p_2\right)$. Fig. \ref{Higgs_decay} shows the Feynman diagram for the decay. 
The decay rate for this process is:
$$
\Gamma=\frac{4\pi}{4 M} \frac{|\bar{M}|^2}{(2 \pi)^2} \frac{\sqrt{E_1^2-m_1^2}}{2M}
$$
with
$$
\begin{aligned}
|\bar{M}|^2 & =2y_{ij}^2 \left(M^2-(m_1+m_2)^2\right)
\end{aligned}
$$

Now for $m_1 + m_2 > M$, the decay width will not be positive real hence no decay contribution will be there if the sum of mass of BSM fields is bigger than Higgs boson mass. One can easily tweak the parameters of the described models to get gear of masses heavier than Higgs mass while surviving the BR, and collider constraints and also satisfying the experimentally observed value for neutrino mass. So, Higgs width will not achieve any new gear contributions and no further constraints from Higgs invisible decay width data will be imposed for gear field masses $> 125 GeV$.

\section{Conclusion}
In this paper, firstly the variants of well-known clockwork suppression models have been explored. It is shown that the underlying mechanism for variants of clockwork is the same for all cases. We found that for some parameter ranges, the variants of clockwork will require a lesser number of sites to produce the same scale as the ordinary clockwork. The analytical expression for 0-mode in some cases of variants of clockwork is found to contain a combinatorial factor which for $0.5 < \left| q \right| < 2$ gives significantly enhanced localization and produces smaller scale than ordinary clockwork and for $\left| q \right| > 2$ differed from a few factors to few orders of magnitude from the clockwork depending on the numbers of gears considered. Also, the non-local clockwork models relax the constraint of $ \left| q \right| > 1$ to get localization, in these models the localization of 0-mode can be achieved even with coupling terms being equal to mass terms. From the Table \ref{tab:clockwork_model_comparison}, it can be seen that the smallest mass scale produced by CW for $q=1$ with n = 40 and Yukawa strength of $O(0.1)$ is $O(10^{-3})$ of the fundamental parameter scale whereas with the same number of sites and with $q = q' = 1$ NNN-CW can produce mass scale of $O(10^{-11})$. As the number of sites increases, the NNN-CW will produce a bigger suppression scale compared to CW since the combinatorial factor will increase. Similarly, CN-CW due to even larger combinatorics factor will produce even larger suppressions. Then, we have mentioned the fine cancellation (precision-prune) mechanism to produce hierarchically small scales from natural order fundamental parameters in dimensional deconstruction perspective. This mechanism has been applied to SM to account for small neutrino mass problems. It can also be extended to account for PMNS mixing of SM active neutrinos along with their masses. The cancellation mechanism is applied in this paper for a specific case of local underlying theory space though one can consider other non-local theory spaces as well.\\
 Finally, phenomenology for all these models was studied using the observable FCNC BR of $\mu \rightarrow e \gamma$ process and some comments were made about BSM effects in colliders and Higgs width/mass. We have laid out some benchmark points which satisfactorily produce neutrino mass as per the current experimental observations of mass squared difference and are also surviving current MEG experiment bounds. These Benchmark Points are within the reach of future MEG experiments and hence will be tested soon. All the BSM Lagrangians discussed in this paper can be straightforwardly extended to Majorana neutrino cases.

\section*{Data Availability Statement}
This manuscript has no associated data.

\section*{Code Availability Statement}
All code used to perform the analysis and generate the results presented in this manuscript is publicly available at:  
\url{https://github.com/AadarshSingh0/Clockwork-Mass-Models}.

%
% BibTeX users please use
% \bibliographystyle{}
% \bibliography{}

\bibliographystyle{unsrt}
\bibliography{bib}

% etc
\section*{Acknowledgement}
The author would like to express his sincere gratitude to Sudhir Vempati for his insightful discussions and guidance throughout the research process. The author would also like to thank G. Kurup for clarifying their work in the same direction. AS thanks CSIR, Govt. of India for SRF fellowship No. 09/0079(15487)/2022-EMR-I. The author further acknowledges various community resources and open-source software tools that were invaluable in the data analysis and visualization aspects of this work, including \textit{Mathematica}, \textit{Python}, and associated scientific computing libraries.
All code used for data generation and analysis in this paper is openly available at:  
\url{https://github.com/AadarshSingh0/Clockwork-Mass-Models},  
and is released under an open-source Creative Commons (CC) license to support transparency, reproducibility, and further research.

\appendix
\section*{Appendix}
\section{Linear Algebra Results}
\label{app:linear_algebra}
The following is a well-known theorem in linear algebra that is used to prove the existence of  0-mode in all variants of CW throughout the paper.\\
\textbf{Theorem :} Let M be an m×n matrix and N be an n×k matrix then rank(MN) $ \leq $ rank(M).\\ 
\textbf{Proof :} Consider N as matrix A and use the fact that the rank of an m×n matrix M is the dimension of the range R(M) of the matrix M with the range given by
$$R(M) = \left\{ y \in \mathbb{R}^m \mid y = Mx \text{ for some } x \in \mathbb{R}^n \right\}
$$
Hence proved. \(\square\) \\
So for a rectangular matrix A of dimension $n \times (n+1)$, the matrix $A^\dagger$$A$ will always have a non-zero dimensional kernel space.

Now, consider a matrix with the rotated left and right basis under Singular value decomposition (SVD) to be $\chi_L$ and $\chi_R$ respectively i.e, 
$$ L = \mathcal{U}\chi_L, \hspace{1cm}
 R = \mathcal{V}\chi_R $$
It's easy to show that for a symmetric matrix, the SVD gives identical left and right modes rotation. For a real symmetric positive definite  matrix M,
\begin{align}
 M^{\dagger} &= M, \nonumber \\ 
 MM^{\dagger}  &= M^{\dagger}M \nonumber \\
 \mathcal{U}^{\dagger}MM^{\dagger}\mathcal{U} & = \mathcal{M}_{diag} \nonumber \\
 \mathcal{U}^{\dagger}M^{\dagger}M\mathcal{U} & = \mathcal{M}_{diag} \nonumber \\
 \mathcal{V}^{\dagger}M^{\dagger}M\mathcal{V} & = \mathcal{M}_{diag}
\end{align}
hence, $ \mathcal{U} $ = $\mathcal{V}$ though $\mathcal{U}$ and $\mathcal{V}$ are unique only when the original matrix is positive definite.

\section{CW and variants Benchmark Points} \label{App:BP}
In this Appendix, the neutrino masses are produced for the neighbouring coupling strength parameter with a value of unity. The local clockwork with $|q| =1$ can never produce hierarchy as the localization of 0-mode is not possible in this case demonstrated in Table \ref{tab:clockwork_model_comparison}. In contrast, due to the localization of the zero mode in non-local variants, the generation of suppressed neutrino masses becomes feasible. Table \ref{tab:bp_points_4} lists the benchmark parameter points required in the NNN-CW framework to produce neutrino masses consistent with current experimental observations, while Table \ref{tab:bp_points_5} presents the corresponding benchmarks for the CN-CW scenario.

\begin{table}[h!]
\caption{BP points for the model with $q_i = q, q'_i = q' \hspace{0.5cm} \forall i \in \{1, 2, \ldots, n \}$ producing neutrino mass in agreement with current experimental data.}
\label{tab:bp_points_4} % Give a unique label
\centering
\begin{tabular}{lllll}  % Changed to five left-aligned columns
\hline\noalign{\smallskip}
\textbf{Hierarchy} & \textbf{n} & \textbf{$q$} & \textbf{$q'$} & \textbf{Yukawa Couplings} \\
\noalign{\smallskip}\hline\noalign{\smallskip}
Inverted & 55 & $-1$ & $-1$ & $ \{y_1, y_2, y_3 \} = \{0.1, 0.101, 0.06 \}$ \\
\noalign{\smallskip}\hline\noalign{\smallskip}
Normal & 55 & $-1$ & $-1$ & $ \{y_1, y_2, y_3 \} = \{0.1, 0.101, 0.13 \}$ \\
\noalign{\smallskip}\hline
\end{tabular}
\end{table}

\begin{table}[h!]
\caption{BP points for the model with $q_i^j = q \hspace{0.2cm} \forall \hspace{0.2cm} i,j$ producing neutrino mass in agreement with experimental data.}
\label{tab:bp_points_5} % Give a unique label
\centering
\begin{tabular}{llll}  % Changed to four left-aligned columns
\hline\noalign{\smallskip}
\textbf{Hierarchy} & \textbf{n} & \textbf{$q$} & \textbf{Yukawa Couplings} \\
\noalign{\smallskip}\hline\noalign{\smallskip}
Inverted & 40 & $-1$ & $ \{y_1, y_2, y_3 \} = \{0.1, 0.105, 0.088 \}$ \\
\noalign{\smallskip}\hline\noalign{\smallskip}
Normal & 40 & $-5$ & $ \{y_1, y_2, y_3 \} = \{0.1, 0.105, 0.12 \}$ \\
\noalign{\smallskip}\hline
\end{tabular}
\end{table}

\section{Higgs Decay Width}
The decay rate for this process is:
$$
d \Gamma=\frac{1}{2 M} \frac{d^3 \Vec{p_1}}{(2 \pi)^3 2 E_1} \frac{d^3 \Vec{p_2}}{(2 \pi)^3 2 E_2}(2 \pi)^4 \delta^4\left(q-p_1-p_2\right)|\bar{M}|^2,
$$
with $|\bar{M}|^2$ as the invariant matrix element squared. Momentum conservation imposes $p_1 \cdot p_2=\left(M^2-m_1^2-m_2^2\right) / 2$ with $p_1^2=m_1^2$ and $p_2^2=m_2^2$. 
The corresponding amplitude $M$ is given by:
$$
\left.M=-i y_{ij} \bar{u}\left(p_1\right) v_{(} p_2\right)
$$
leading to:
$$
\begin{aligned}
|\bar{M}|^2 & =y_{ij}^2\left(\operatorname{Tr}\left[\not p_1 \not p_2\right]-m_1m_2 \operatorname{Tr}[1]\right) \\
& = 4p_1.p_2-4m_1m_2 \\
& =2y_{ij}^2 \left(M^2-(m_1+m_2)^2\right) .
\end{aligned}
$$
Since the amplitude square $|M|^2$ is momentum-independent it can be taken outside the integration. Using $d^3 \Vec{p_2} / 2 E_2=d^4 p_2 \delta^{+}\left(p_2^2-m_2^2\right)$ and carrying out the $d^4 p_2$ integration it comes out
$$
d \Gamma=\frac{1}{2 M} \frac{|\bar{M}|^2}{(2 \pi)^2} \int \frac{d^3 \Vec{p_1}}{2 E_1} \delta^{+}\left(\left(q-p_1\right)^2-m_2^2\right) .
$$
In the rest frame of the Higgs boson, $q=(M, 0,0,0)$, hence the argument of the $\delta^{+}$ function reduces to \\
$\left(M^2-2 M E_1+m_1^2-m_2^2\right)$ . Now, integrating the expression over $\Vec{p1}$ to get the decay width as:
$$
d \Gamma=\frac{1}{2 M} \frac{|\bar{M}|^2}{(2 \pi)^2} \int \frac{\Vec{p_1^2} d \Vec{p_1} d\phi d \theta sin\theta}{2 E_1} \delta^{+}\left(M^2-2ME_1+m_1^2-m_2^2\right) .
$$
using the fact that $\Vec{p_1} d \Vec{p_1}$ = $E_1 d E_1$,
\begin{equation}
\scalebox{0.9}{$
d\Gamma = \frac{1}{4 M} \frac{|\bar{M}|^2}{(2\pi)^2} 
\int \frac{\vec{p}_1 E_1\, dE_1\, d\phi\, d\theta\, \sin\theta}{E_1} 
\delta^{+}\left(M^2 - 2ME_1 + m_1^2 - m_2^2\right)
$}
\end{equation}
This integral is non-zero for the value of $E_1$ that will lead to the Dirac-Delta function with 0 as the argument i.e,
\begin{align}
 M^2&-2ME_1 +m_1^2-m_2^2 = 0 \nonumber \\
& E_1 = \frac{M^2+m_1^2-m_2^2}{2M} 
\end{align}
$$
\Gamma=\frac{4\pi}{4 M} \frac{|\bar{M}|^2}{(2 \pi)^2} \frac{\sqrt{E_1^2-m_1^2}}{2M}
$$

\section{One Loop Calculations} \label{app:one-loop}
The amplitude for 1-loop with fermions of different masses is given by
\begin{align} 
\scalebox{0.98}{$ i \Pi_{1 L}=-N_f(-iy_{ij})^2\operatorname{Tr}\left[\int \frac{d^4 k}{(2 \pi)^4} \frac{i(\not{k}+m_i)}{\left[k^2-m_i^2\right]} \frac{i(\not{k}+\not{q}+m_j)}{\left[(k+q)^2-m_j^2\right]}\right]$} 
\end{align}
$N_f$ is the number of flavours. In the limit when both fermions in the loop are the same, $m_i$ = $m_j$ = m and $y_{ij}$ = $y_{ii}$. Using dimensional regularization, we get the familiar result
$$
\Pi\left(q^2\right)= N_f\frac{12 y_{ii}^2}{(4 \pi )^2} \mu^{2 \epsilon} \int_0^1 d x \Delta^2\left(\frac{1}{\hat{\epsilon}}+\ln \left(\frac{\Delta^2}{\mu^2}\right)-\frac{1}{3}\right)
$$
where,
$$
\int_0^1 d x \Delta^2=\int_0^1 d x\left(-q^2 x(1-x)+m^2\right)=m^2-\frac{q^2}{6}
$$
Thus, we can write the following expression for the amplitude:
\begin{align}
\Pi\left(q^2\right)= & N_f\frac{12 y_{ii}^2}{(4 \pi )^2} \mu^{2 \epsilon}\Big[\frac{1}{\hat{\epsilon}}\left(m^2-\frac{q^2}{6}\right)+\frac{q^2}{18}-\frac{m^2}{3}+ \nonumber \\ & \int_0^1 d x \Delta^2 \ln \left(\frac{\Delta^2}{\mu^2}\right)\Big]
\end{align}
Using the $\overline{M S}$ scheme we obtain the radiative mass correction to Higgs as:
\begin{align}
\Pi_R(q, \mu)= N_f\frac{12 y_{ii}^2}{(4 \pi 
)^2}\left(\frac{q^2}{18}-\frac{m^2}{3}+\int_0^1 d x \Delta^2 \ln \left(\frac{\Delta^2}{\mu^2}\right)\right)
\end{align}
Now, consider the scenario with two different fermions coupled to Higgs, $m_i$ $\neq$ $m_j$ and coupling with Higgs is $y_{ij}$. The amplitude is 
\begin{align}
\scalebox{0.99}{$i \Pi_{1 L}=-N_f(-iy_{ij})^2\operatorname{Tr}\left[\int \frac{d^4 k}{(2 \pi)^4} \frac{i(\not{k}+m_i)}{\left[k^2-m_i^2\right]} \frac{i(\not{k}+\not{q}+m_j)}{\left[(k+q)^2-m_j^2\right]}\right]
$}
\end{align}
\begin{align}
=  -N_fi^4 y^2_{ij}\int \frac{d^4 k}{(2 \pi)^4} \operatorname{Tr}\left[\frac{(\not{k}+m_i)}{\left[k^2-m_i^2\right]} \frac{(\not{k}+\not{q}+m_j)}{\left[(k+q)^2-m_j^2\right]}\right] \label{51}
\end{align}
using, $$Tr\{(\not{k}+m_i)(\not{k+q}+m_j)\} = 4(k.q + k^2 + m_im_j)$$
plugging this in \eqref{51} and using Feynman parametrization of the propagators i.e,
$$
\frac{1}{A B}=\int_0^1 d x \frac{1}{[A x+B(1-x)]^2}
$$
Taking $A=(k+q)^2-m_i^2$ and $B=k^2-m_j^2$ we get:
\begin{align}
& Ax + B(1-x) = x[(k+q)^2-m_j^2] + (k^2-m_i^2)(1-x) \nonumber \\ &
= (k^2 + q^2 + 2qk - m_j^2)x + k^2 - m_i^2 - k^2x + m_i^2x \nonumber \\ &
= k^2 - m_i^2 +q^2x+2qkx -m_j^2x +m_i^2x \pm q^2x^2 \nonumber \\ &
= (k+qx)^2 +q^2x - q^2x^2 - m_j^2x + m_i^2x -m_i^2 \nonumber \\ &
= (k+qx)^2 + q^2x(1-x) -m_i^2 +x(m_i^2-m_j^2) \nonumber \\ &
= (k+qx)^2 - \Delta'^2
\end{align}
with $\Delta'^2$ = -$(q^2x(1-x) -m_i^2 +x(m_i^2-m_j^2))$, we see that in the limit $m_i$ = $m_j$, $\Delta'^2$ = $\Delta^2$.
$$
\frac{1}{\left(k^2-m_i^2\right)\left[(k+q)^2-m_j^2\right]}=\int_0^1 d x \frac{1}{\left[(k+q x)^2-\Delta'^2\right]^2}
$$
plugging this parametrization in \eqref{51}, We obtain the following amplitude expression:
$$
i \Pi\left(q^2\right)=-4N_f y_{ij}^2 \int_0^1 d x \int \frac{d^4 k}{(2 \pi)^4} \frac{k^2+m_im_j+k q}{\left[(k+q x)^2-\Delta'^2\right]^2}
$$
After performing the variable shift $k \rightarrow k+x q$ $\equiv$ $l$, the numerator becomes:
\begin{align}
    k^2+m_im_j+k q &= (l-qx).q +(l-qx)^2+m_im_j \nonumber \\ &
    = l.q -q^2x +l^2 +q^2x^2 -2l.qx +m_im_j
\end{align}
eliminating all the linear terms in $l^\mu$ and redefining $l$ as $k$, we obtain:
\begin{align}
i \Pi\left(q^2\right)=-4 N_f y^2_{ij} \int_0^1 d x \int \frac{d^4 k}{(2 \pi)^4} \frac{k^2+m_im_j-q^2x(1-x)}{\left[k^2-\Delta'^2\right]^2} \label{54}
\end{align}
Next, we use Dimensional Regularization to compute the integrals of the following type
\begin{align}
    I_1 = \int \frac{d^dk}{(2\pi)^d} \frac{k^2}{(k^2+\Delta')^2} = \frac{1}{(4\pi)^\frac{d}{2}}\frac{d}{2}\frac{\Gamma(2-\frac{d}{2}-1)}{\Gamma(2)}\Big(\frac{1}{\Delta'}\Big)^{2-\frac{d}{2}-1}
\end{align}
and
\begin{align}
    I_2 = \int \frac{d^dk}{(2\pi)^d} \frac{1}{(k^2+\Delta')^2} = \frac{1}{(4\pi)^\frac{d}{2}}\frac{d}{2}\frac{\Gamma(2-\frac{d}{2})}{\Gamma(2)}\Big(\frac{1}{\Delta'}\Big)^{2-\frac{d}{2}}
\end{align}
using dim. reg. in \eqref{54} about $d$ = $4 - \epsilon$, and expanding up to $O(\epsilon)$ we get:\\
{\small
\begin{align}
& \Pi\left(q^2\right)=- 4N_fy^2_{ij} \int_0^1 dx \Bigg[\frac{\Delta'^2}{4\pi^2\epsilon} + \frac{\Delta'^2(2log(\mu^2)-2\gamma_E+1)}{16\pi^2}   \nonumber \\ & +\frac{\Delta'^2(log(16\pi^2)-2log(-\Delta'^2))}{16\pi^2}
 +(m_im_j -q^2x(1-x))\Big\{ \frac{1}{8\pi^2\epsilon} - \nonumber \\ & \frac{\gamma_E+log(-\Delta'^2) - log(4\pi)-log(\mu^2)}{16\pi^2} \Big\} \Bigg] +O(\epsilon)
\end{align}
}
using the renormalization scheme and putting all the divergences in the counterterms, we get\\
{\small
\begin{align}
& \Pi_R\left(q^2, \mu \right)=- 4N_fy^2_{ij} \int_0^1 dx \Bigg[ \frac{\Delta'^2(1-2log(\frac{-\Delta'^2}{\mu^2}))}{16\pi^2} + \nonumber \\ &
 (m_im_j -q^2x(1-x))\Big\{  - \frac{log(\frac{-\Delta'^2}{\mu^2})}{16\pi^2} \Big\} \Bigg] +O(\epsilon) \nonumber \\ &
 = - \frac{4 N_f y^2_{ij}}{16\pi^2} \Bigg[ \frac{m_{i}^2}{2}+\frac{1}{2} \Big(m_{j}^2-q^2\Big)+\frac{q^2}{3} -\frac{1}{18q^4} \times \nonumber \\ & \Big\{ 6 \Big(m_{i}^4-2 m_{i}^2 \left(m_{j}^2+q^2\right)+\left(m_{j}^2-q^2\right)^2\Big)^{3/2} \times \nonumber \\ & \tanh ^{-1}\Big(\frac{m_{i}^2-m_{j}^2-q^2}{\sqrt{m_{i}^4-2 m_{i}^2 \left(m_{j}^2+q^2\right)+\left(m_{j}^2-q^2\right)^2}}\Big)-\nonumber \\ &6 \Big(m_{i}^4-2 m_{i}^2 \left(m_{j}^2+q^2\Big)+\left(m_{j}^2-q^2\right)^2\right)^{3/2} \nonumber \\ & \times \tanh ^{-1}\Bigg(\frac{m_{i}^2-m_{j}^2+q^2}{\sqrt{m_{i}^4-2 m_{i}^2 \left(m_{j}^2+q^2\right)+\left(m_{j}^2-q^2\right)^2}}\Bigg)+\nonumber \\ &2 q^2 \Bigg(3 m_{i}^4-3 q^2 \left(-3 m_{i}^2-3 m_{j}^2+q^2\right) \times \log \left(-\frac{m_j^2}{\mu ^2}\right)\nonumber \\ &-6 m_{i}^2 \left(m_{j}^2+2 q^2\right)+3 m_{j}^4-12 m_{j}^2 q^2+5 q^4\Bigg)-3 \log \left(m_{i}^2\right) \nonumber \\ &\Bigg(m_{i}^6-3 m_{i}^4 \left(m_{j}^2+q^2\right) +3 m_{i}^2 \left(m_{j}^4-q^4\right)-\left(m_{j}^2-q^2\right)^3\Bigg)\nonumber \\ &+3 \log \left(m_{j}^2\right) \Big(m_{i}^6-3 m_{i}^4 \left(m_{j}^2+q^2\right)+3 m_{i}^2 \left(m_{j}^4-q^4\right)- \nonumber \\ & -36 m_{i} m_{j} q^2-6 m_{j}^4+6 m_{j}^2 q^2+5 q^4\Big)-\nonumber \\ &6 \sqrt{m_{i}^4-2 m_{i}^2 \left(\text{mj}^2+q^2\right)+\left(\text{mj}^2-q^2\right)^2} \Big(2 m_{i}^4- m_{i}^2   \left(4 m_{j}^2+q^2\right)\nonumber \\ &+6 m_{i} m_{j} q^2+2 m_{j}^4-m_{j}^2 q^2-q^4\Big)\nonumber \\ & \tanh ^{-1}\left(\frac{m_{i}^2-m_{j}^2-q^2}{\sqrt{m_{i}^4-2 m_{i}^2 \left(\text{mj}^2+q^2\right)+\left(\text{mj}^2-q^2\right)^2}}\right) \nonumber \\ & +6 \left(2 m_{i}^4-m_{i}^2 \left(4 m_{j}^2+q^2\right)+6 m_{i} m_{j} q^2+2 m_{j}^4-m_{j}^2 q^2-q^4\right) \nonumber \\ &\sqrt{m_{i}^4-2 m_{i}^2 \left(\text{mj}^2+q^2\right)+\left(\text{mj}^2-q^2\right)^2} \nonumber \\ & \tanh ^{-1}\left(\frac{m_{i}^2-m_{j}^2+q^2}{\sqrt{m_{i}^4-2 m_{i}^2 \left(\text{mj}^2+q^2\right)+\left(\text{mj}^2-q^2\right)^2}}\right)\nonumber \\ &-3 \log \left(m_{j}^2\right) \Big(2 m_{i}^6-3 m_{i}^4 \left(2 m_{j}^2+q^2\right) \nonumber \\ & +6 m_{i}^3 m_{j} q^2+6 m_{i}^2 m_{j}^4+m_{i} \left(6 m_{j} q^4-6 m_{j}^3 q^2\right)-2 m_{j}^6+3 m_{j}^4 q^2-q^6\Big)\nonumber \\ &+3 \log \left(m_{i}^2\right) \Big(2 m_{i}^6-3 m_{i}^4  \left(2 m_{j}^2+q^2\right)+6 m_{i}^3 m_{j} q^2+6 m_{i}^2 m_{j}^4+\nonumber \\ &m_{i} \left(6 m_{j} q^4-6 m_{j}^3 q^2\right)-2 m_{j}^6+3 m_{j}^4 q^2-q^6\Big) \Big\} \Bigg]
\end{align}    
}
the energy independent contribution to Higgs's mass is proportional to coupling$\times$mass of fermions.
\begin{align}
    \delta m_h^2 \propto (y^2_{ij})(m_i^2 + m_j^2)
\end{align}

\end{document}